\documentclass[12pt]{iopart-AG}
\usepackage{setstack}
\usepackage{graphicx,amssymb,amsthm,epsfig,amsbsy,amsfonts,mathrsfs}
\usepackage[english,french]{babel}

\usepackage{ulem}

\usepackage{url,hyperref}

\usepackage{cite}

\usepackage[usenames,dvipsnames]{xcolor}

\definecolor{Cblue}{HTML}{045FB4}
\definecolor{Cred}{HTML}{DF0101}

\hypersetup{
  colorlinks=true, %set true if you want colored links
  linktoc=true,     %set to all if you want both sections and
  linkcolor=black,
  citecolor=Cblue, 
  urlcolor=Cblue
}

\usepackage{hhline}

\renewcommand{\em}{\it}

\renewcommand{\leq}{\leqslant}

\newcommand{\ket}[1]{|\kern.3ex#1\kern.3ex\rangle}
\newcommand{\bra}[1]{\langle\kern.3ex #1 \kern.3ex|}

\def\dd{{\rm d}}                  % la differenciation
\def\D{{\mathcal{D}}}                 % D integrale fonctionnelle

\def\Xint#1{\mathchoice
  {\XXint\displaystyle\textstyle{#1}}%
  {\XXint\textstyle\scriptstyle{#1}}%
  {\XXint\scriptstyle\scriptscriptstyle{#1}}%
  {\XXint\scriptscriptstyle\scriptscriptstyle{#1}}%
  \!\int}
\def\XXint#1#2#3{{\setbox0=\hbox{$#1{#2#3}{\int}$}
    \vcenter{\hbox{$#2#3$}}\kern-.5\wd0}}

\def\dashint{\Xint-}

\newcommand{\abs}[1]{\left| #1 \right|}

\usepackage{etoolbox}
\makeatletter
\def\@mkboth#1#2{}
\newlength\appendixwidth
\preto\appendix{\addtocontents{toc}{\protect\patchl@section}}
\newcommand{\patchl@section}{%
  \settowidth{\appendixwidth}{\textbf{Appendix }}%
  \addtolength{\appendixwidth}{1.5em}%
  \patchcmd{\l@section}{1.5em}{\appendixwidth}{}{\ddt}%
}
\makeatother

\newcommand{\moy}[1]{\left\langle #1\right\rangle}

\def\Pnk{P_{N,\kappa}}

\def\e{\mathrm{e}}

\def\O{\mathcal{O}}

\def\rhoL{\rho_{\mathrm{L}}}
\def\rhoR{\rho_{\mathrm{R}}}

\begin{document}

\selectlanguage{english}

%%%%%%%%%%%%%%%%%%%%%%%%%%%%%%%%%%%%%%%%%%%%%%%%%%%%%%%%%%%%%%%%%%%%%%%%%%%%% 
\renewcommand{\labelitemi}{$\bullet$}
\renewcommand{\labelitemii}{$\star$}
%%%%%%%%%%%%%%%%%%%%%%%%%%%%%%%%%%%%%%%%%%%%%%%%%%%%%%%%%%%%%%%%%%%%%%%%%%%%% 

\title{General truncated linear statistics for the top eigenvalues of random matrices}

\author{Aur\'elien Grabsch}
\address{Sorbonne Universit\'e, CNRS, Laboratoire de Physique Th\'eorique de la Mati\`ere Condens\'ee (LPTMC), 4 Place Jussieu, 75005 Paris, France}

\date{today}

\begin{abstract}
  Invariant ensemble, which are characterised by the joint
  distribution of eigenvalues $P(\lambda_1,\ldots,\lambda_N)$, play a
  central role in random matrix theory.  We consider the truncated
  linear statistics $L_K = \sum_{n=1}^K f(\lambda_n)$ with
  $1 \leq K \leq N$, $\lambda_1 > \lambda_2 > \cdots > \lambda_N$ and
  $f$ a given function. This quantity has been studied recently in the
  case where the function $f$ is monotonous. Here, we consider the
  general case, where this function can be non-monotonous. Motivated
  by the physics of cold atoms, we study the example
  $f(\lambda)=\lambda^2$ in the Gaussian ensembles of random matrix
  theory. Using the Coulomb gas method, we obtain the distribution of
  the truncated linear statistics, in the limit $N \to \infty$ and
  $K \to \infty$, with $\kappa = K/N$ fixed. We show that the
  distribution presents two essential singularities, which arise from
  two infinite order phase transitions for the underlying Coulomb
  gas. We further argue that this mechanism is universal, as it
  depends neither on the choice of the ensemble, nor on the function
  $f$.
\end{abstract}

\maketitle

\vspace{10pt}
\noindent\rule{\textwidth}{1pt}
\tableofcontents
\noindent\rule{\textwidth}{1pt}
\vspace{10pt}

\hypersetup{
    linkcolor=Cred
  }

% \pacs{05.60.Gg ; 03.65.Nk ; 05.45.Mt ; 72.15.Rn}

% \pacs{73.20.Fz}{Weak or Anderson localisation}

% 02.50.-r Probability theory, stochastic processes, and statistics
% 02.50.Cw 	Probability theory 
% 02.50.Ey 	Stochastic processes 
% 03.65.Nk    Scattering theory 
% 05.10.Gg 	Stochastic analysis methods (Fokker-Planck, Langevin, etc.) 
% 05.40.-a 	Fluctuation phenomena, random processes, noise, and
% 05.40.Jc Brownian motion
% 05.45.Mt    Quantum chaos ; semiclassical methods 
% 05.60.Gg    Quantum transport
% 05.60.-k 	Transport processes
% 05.70.Np  Interface and surface thermodynamics
% 72.   Electronic transport in condensed matter
% 72.10.-d   Theory of electronic transport; scattering mechanisms
% 72.10.Bg   General formulation of transport theory 
% 72.15.Rn Localization effects (Anderson or weak localization)

% 73.   Electronic structure and electrical properties of surfaces, interfaces,
% thin films, and low-dimensional structures 
% 73.23.-b     Electronic transport in mesoscopic systems 
% 73.20.Fz Weak or Anderson localization

%%%%%%%%%%%%%%%%%%%%%%%%%%%%%%%%%%%%%%%%%%%%%%%%%%%%%%%%%%%%%%%%%%%%%%%%%%%%%%%%%%%%%%%%%% 
%%%%%%%%%%%%%%%%%%%%%%%%%%%%%%%%%%%%%%%%%%%%%%%%%%%%%%%%%%%%%%%%%%%%%%%%%%%%%%%%%%%%%%%%%% 

\section{Introduction}
\label{sec:Introduction}

Random matrix theory has first been introduced in physics by Wigner
and Dyson in the 1950s to model the atomic nucleus\cite{Wig51}. It has
now been successfully applied in various domains of physics, such as
electronic quantum
transport~\cite{Bee97,GuhMulWei98,AleBroGla02,MelKum04,Bro95,MelBar99,SomWieSav07,VivMajBoh08,KhoSavSom09,VivMajBoh10,VivViv08,GraTex15,CunFacViv15},
quantum
information~\cite{Pag93,FacMarParPacS08,PasFacParPacSca10,NadMaj10,NadMajVer11,FacFloParPasYua13},
statistical physics of fluctuating interfaces~\cite{NadMaj09,Nad11} or
cold atoms\cite{MarMajSchViv16,DeaLeDMajSch19}.

Invariant ensembles play a prominent role in random matrix theory. They correspond to distributions of matrices which are invariant under changes of basis. Consequently, the eigenvalues and eigenvectors are statistically independent, and one can focus on the eigenvalues only. In the most famous invariant ensembles, the joint distribution of the eigenvalues takes the form\cite{Meh04,For10,AkeBaiDiFra11}
\begin{equation}
  \label{eq:DefInvEns}
  P(\lambda_1,\ldots, \lambda_N) \propto
  \prod_{i<j} \abs{\lambda_i - \lambda_j}^\beta
  \prod_{n=1}^N \e^{- \frac{\beta N}{2} V(\lambda_n)}
  \:,
\end{equation}
where $\beta$ is the Dyson index which classifies real ($\beta=1$),
complex ($\beta=2$) and quaternionic ($\beta=4$) matrices, and
$V(\lambda)$ depends on the specific choice of the ensemble (see
Table~\ref{tab:Ensembles} for explicit expressions for the standard
ensembles). The prefactor in the exponential has been chosen such that
the eigenvalues do not scale with $N$, i.e. $\lambda_n = \O(N^0)$ for
$N \to \infty$, and $\beta$ has been introduced for convenience.

\begin{table}
  \centering
  \begin{tabular}{|c|c|c|}
    \hline
    Ensemble & Domain & $V(\lambda)$
    \\
    \hhline{|=|=|=|}
    Gaussian & $\lambda \in \mathbb{R}$ & $\lambda^2$
    \\
    \hline
    Laguerre & $\lambda \in \mathbb{R}^+$ & $\lambda - \nu \ln \lambda$
    \\
    \hline
    Jacobi & $\lambda \in [0,1]$ & $-a \ln \lambda - b \ln (1-\lambda)$
    \\
    \hline
    Cauchy & $\lambda \in \mathbb{R}$ & $a \ln (1+\lambda^2)$
    \\
    \hline
  \end{tabular}
  \caption{Example of invariant ensembles of random matrices. The joint distribution of eigenvalues is given by Eq.~(\ref{eq:DefInvEns}). }
  \label{tab:Ensembles}
\end{table}

Many applications of random matrices, and in particular of invariant ensembles, rely on the study of linear statistics of the eigenvalues, which are quantities of the form
\begin{equation}
  \label{def:LinStat}
  L = \sum_{n=1}^N f(\lambda_n)
  \:,
\end{equation}
where $f$ can be any given function (not necessarily linear). Many
relevant quantities can indeed be expressed in this form, such as the
number of eigenvalues in a given
domain\cite{MajNadScaViv09,MajNadScaViv11,MajViv12,MarMajSchViv14,MarMajSchViv14a,MarMajSchViv16},
the conductance and shot noise of a quantum
dot\cite{Bee97,SomWieSav07,KhoSavSom09}, or the mutual information in
MIMO\footnote{Multiple-input, multiple-output}
channels\cite{KazMerMouCai11,KarMouViv14}.  Several methods have been
developed to analyse the statistical properties of linear
statistics. The typical fluctuations can be studied using orthogonal
polynomials, or Selberg's integral. In particular, Dyson and Mehta
have obtained a general formula for the variance of linear statistics
in the Gaussian ensembles\cite{DysMeh63}, and other formulae can be
found in the literature\cite{Bee93,Bee93b,Bee94,BasTra93,JanFor94}.
More recently, the question of atypical fluctuations, associated to
rare events, has been considered\cite{DeaMaj06,VivMajBoh07}. The best
suited tool to address this question is the Coulomb gas
method\cite{Dys62I-III,AroGui97,AroZei98}.  The joint distribution of
eigenvalues~(\ref{eq:DefInvEns}) is interpreted as a Gibbs weight for
a gas of particles at positions $\lambda_n$, which repel
logarithmically. The determination of the distribution of the linear
statistics reduces to the determination of the configuration of the
particles which minimises the energy of the gas under the
constraint~(\ref{def:LinStat}) with fixed $L$. An interesting feature
of this approach is the possibility of phase transitions in the
Coulomb gas (changes of the shape of the density of eigenvalues),
which manifest themselves as non-analyticities in the distribution of
the linear statistics in the limit $N \to \infty$. For an overview of
the different types of transition, see the table at the end of
Ref.\cite{GraMajTex16}.

Recently, several extensions of these linear statistics have been
considered, in which the summation in~(\ref{def:LinStat}) does not run
over all the eigenvalues.  One of these extensions is based on the
work of Bohigas and Pato~\cite{BohPat06}, who considered the effect of
randomly removing each eigenvalue with a given probability. This
situation is described by the so-called \textit{thinned ensembles},
which have been the focus of several works over the last
years~\cite{BerDui17,ChaCla17,Lamb16}, studying in particular linear
statistics (see also~\cite{GraMajTex17} for a related problem).

Alternatively, one can choose deterministically a subset of the
eigenvalues\footnote{A well-known duality between ensembles of random
  matrices is obtained by selecting a given subset of eigenvalue. More
  precisely, the set of every second eigenvalues
  $\{ \lambda_{2n} \}_{n=1,\ldots,N}$ of a matrix of size
  $(2N+1)\times(2N+1)$ in the Gaussian Orthogonal Ensemble ($\beta=1$)
  is distributed as the eigenvalues of a $N\times N$ matrix from the
  Gaussian Symplectic Ensemble ($\beta=4$). Similar duality relations
  exist for other ensembles of random matrices~\cite{For10}.}
$\{ \lambda_n \}$, and compute the associated linear statistics. This
situation has been first considered in\cite{GraMajTex16}, where only a
given number $K \leq N$ of the largest eigenvalues were selected,
leading to consider the \textit{truncated linear statistics}
\begin{equation}
  \label{def:TruncLinStat}
  L_K = \sum_{n=1}^K f(\lambda_n)
  \:,
  \quad
  \lambda_1 > \lambda_2 > \cdots > \lambda_N
  \:.
\end{equation}
This situation occurs naturally in various contexts, for instance in
principal component analysis, where one focuses on a given number of
the largest eigenvalues since they contain the most relevant
information\cite{Smi02,MajViv12}. The truncated linear
statistics~(\ref{def:TruncLinStat}) interpolates between the usual
linear statistics~(\ref{def:LinStat}) for $K=N$, and the largest
eigenvalue only for $K=1$, which is also a widely studied
quantity~\cite{TraWid94,TraWid96,Joh07,DeaMaj06,VivMajBoh07,DeaMaj08,BorEynMajNad11,MajSch14}. The
statistical properties of such truncated linear statistics have been
studied in Ref.\cite{GraMajTex16} in the bulk regime $N \to \infty$
and $K \to \infty$ with $\kappa = K/N$ fixed, using the Coulomb gas
method. It was shown that the distribution of $L_K$ displays a
singular behaviour at its typical value, which originates from an
infinite order phase transition in the underlying Coulomb gas. This
behaviour is universal, meaning that it neither depends on the choice
of the ensemble, nor on the choice of the function $f$, provided that
it is \textit{monotonous}. This problem has also been considered in
the edge regime $N \gg K \gg 1$, but again for a class of monotonous
functions\cite{KraLeD19}. Finally, truncated linear statistics have
been studied for the one dimensional plasma in
Ref.~\cite{FlaMajSch21}, which is also a one dimensional gas of
particles, but with linear repulsion, and again for a monotonous
function $f$.

The aim of this paper is to consider the general case where the
function $f$ can be non-monotonous. This extension is crucial to study
various important observables, such as the entanglement entropy which
corresponds to $f(\lambda) = -\lambda \ln \lambda$. Our goal is to
determine the distribution of the rescaled truncated linear
statistics~(\ref{def:TruncLinStat})
\begin{eqnarray}
  \label{eq:DistrTruncLS}
  \Pnk &\left( s = \frac{L_K}{N} \right)
         = N! \times\\
  \nonumber
       &\int \dd \lambda_1 \int^{\lambda_1} \dd \lambda_2 \cdots \int^{\lambda_{N-1}} \dd \lambda_N
         \: 
  P(\lambda_1,\ldots,\lambda_N)
  \: \delta \left(
  s - \frac{1}{N} \sum_{n=1}^K f(\lambda_n)
    \right)
\end{eqnarray}
in the bulk regime $N \to \infty$ and $K \to \infty$ with
$\kappa = K/N$ fixed. Although we will argue that our results are
general, we will mostly focus on a specific example in order to make
the analysis more concrete. We will consider the simplest
non-monotonous function $f(\lambda) = \lambda^2$, and work in the
Gaussian ensembles (see Table~\ref{tab:Ensembles}). Besides being the
most elementary example, this situation is also motivated by the
physics of cold atoms: the eigenvalues $\{ \lambda_n \}$ in the
Gaussian ensemble with $\beta=2$ can be interpreted as the positions
of $N$ spinless fermions placed in a one-dimensional harmonic trap at
zero temperature.  The corresponding truncated linear
statistics~(\ref{def:TruncLinStat}) is then the potential energy
carried by the $K$ rightmost fermions.

\subsection{Main results}

In the limit $N \to \infty$, with $\kappa = \frac{K}{N}$ fixed, the
distribution of the truncated linear
statistics~(\ref{def:TruncLinStat}) with $f(\lambda)=\lambda^2$ in the
Gaussian ensembles ($V(\lambda)=\lambda^2$) takes the
form\footnote{Throughout the paper, the notation
  $P_{N,\kappa}(s) \underset{N \to \infty}{\sim} \exp[- \frac{\beta
    N^2}{2} \Phi_\kappa(s)]$ must be understood as
  $\lim_{N \to \infty} \frac{2}{\beta N^2} \ln P_{N,\kappa}(s) = -
  \Phi_\kappa(s)$.}
\begin{equation}
  P_{N,\kappa} \left(
    s = \frac{L_K}{N}
  \right)
  \underset{N \to \infty}{\sim}
  \exp \left[
    - \frac{\beta N^2}{2} \Phi_\kappa(s)
  \right]
  \:,
\end{equation}
where we have introduced the large deviation function $\Phi_\kappa$,
which has the following behaviours:
\begin{equation}
  \Phi_\kappa(s) \simeq
  \left\lbrace
    \begin{array}{ll}
      \displaystyle
      - \frac{\kappa^2}{2} \ln s
      & \text{for } s \to 0 \:,
      \\[0.4cm]
      \displaystyle
      \frac{(s-s_0(\kappa))^2}{2 F(c_0(\kappa))}
      & \text{for } s \to s_0(\kappa) \:,
      \\[0.4cm]
      \displaystyle
      s - \frac{\kappa(2-\kappa)}{2} \ln s
      & \text{for } s \to \infty \:,
    \end{array}
  \right.
\end{equation}
where $s_0(\kappa)$ is the typical value of the truncated linear
statistics $s$, given by Eq.~(\ref{eq:optsC}) below, in terms of
$c_0(\kappa)$~(\ref{eq:optkC}). The function $F(c_0)$ controls the
variance of $s$, and is given explicitly by~(\ref{eq:ExprFc0}). The
function $F$, and thus the variance of $s$, displays a surprising
non-monotonous behaviour as a function of the fraction $\kappa$, as
illustrated in Fig.~\ref{fig:PltVar} below.

\begin{figure}
  \centering
  \includegraphics[width=0.6\textwidth]{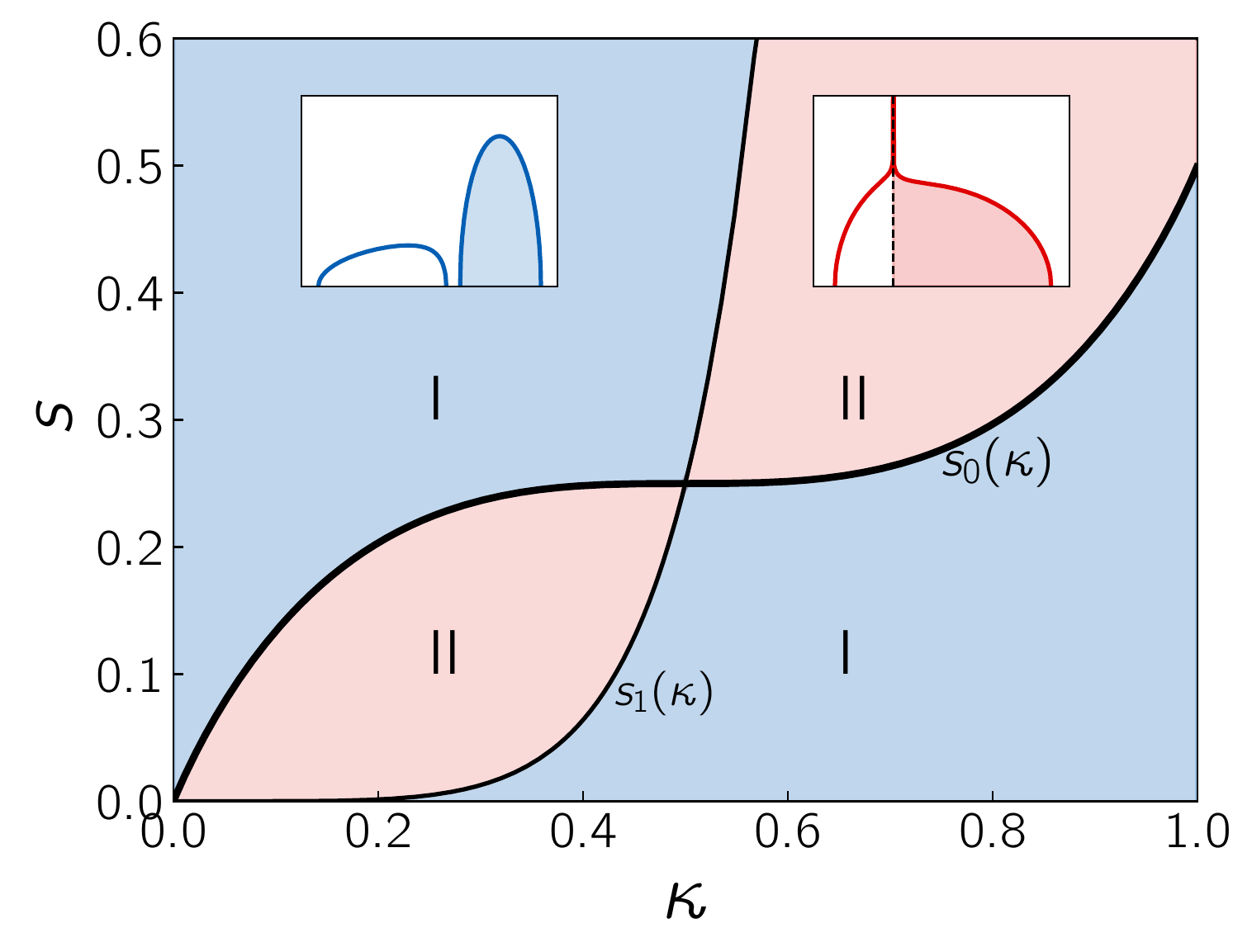}
  \caption{Phase diagram of the Coulomb gas for the truncated linear
    statistics~(\ref{def:TruncLinStat}) with $f(\lambda)=\lambda^2$ in
    the Gaussian ensembles ($V(\lambda) = \lambda^2$). The insets
    represent the shape of the optimal density of eigenvalues in the
    corresponding phases.}
  \label{fig:PhDiag}
\end{figure}

This specific form of the distribution arises from a universal
mechanism for the underlying Coulomb gas. Indeed, in the limit
$N \to \infty$, the distribution of the truncated linear statistics is
dominated by one optimal configuration of the eigenvalues (or charges
of the Coulomb gas). This optimal configuration is determined by the
two parameters: the fraction $\kappa = K/N$ of eigenvalues under
consideration, and $s$ which controls the constraint
in~(\ref{eq:DistrTruncLS}). The corresponding phase diagram is shown
in Fig.~\ref{fig:PhDiag}.

For a fixed value of $\kappa$, the parameter $s$ drives two
consecutive phase transitions for the Coulomb gas, corresponding to a
change in the optimal configuration of eigenvalues which
dominates~(\ref{eq:DistrTruncLS}). The first phase (Phase~I)
corresponds to an optimal density of eigenvalues supported on two
disjoint supports. For instance, for $\kappa< \frac{1}{2}$, it
corresponds to the region $s > s_0(\kappa)$. As $s$ decreases, the gap
between the two supports reduces, until it completely closes when
$s=s_0(\kappa)$. As $s$ is further decreased, entering Phase~II, a
logarithmic divergence emerges at the point where the two interval
have merged. Up to now, this scenario is identical to the one
described in Ref.\cite{GraMajTex16} in the case of monotonous
functions $f$.

The specificity of the non-monotonous situation now appears, as $s$ is
further decreased. The location of the logarithmic divergence is
displaced, until it reaches the origin (or more generally, the point
where $f'(x)=0$). At this point, the divergence vanishes, but the
density still has a logarithmic singularity:
$\rho(x) \underset{x\to 0}{\simeq} \rho(0) - \alpha x \ln \abs{x}$,
with a constant $\alpha$. We denote $s_1(\kappa)$ the corresponding
value of $s$. It it given parametrically by
Eqs.~(\ref{eq:S1K1}-\ref{eq:S1K3}). If $s$ is further decreased, the
density splits at the origin into two supports, re-entering
Phase~I. For $\kappa > \frac{1}{2}$, the order of these transitions
is inverted.

We further show that the energy of the Coulomb gas, corresponding to
the large deviation function $\Phi_\kappa$, has an essential
singularity at each phase transition, $s=s_0(\kappa)$ and
$s=s_1(\kappa)$. We thus speak of \textit{infinite order} phase
transitions. The two transition lines intersect at the specific value
$\kappa = \frac{1}{2}$. At this point, there is no longer a phase
transition in the Coulomb gas, as it remains in Phase~I both for
$s>s_0(\kappa)$ and $s<s_0(\kappa)$.

\bigskip

This scenario is not restricted to the example considered here. We
argue in Section~\ref{sec:Universality} (and~\ref{app:InfOrderTr})
that this scenario is universal: it holds for any matrix ensemble
(choice of $V$) and any linear statistics ($f$), at least in the
vicinity of the line $s_0(\kappa)$ which is the typical value taken by
the truncated linear statistics~(\ref{def:TruncLinStat}). Near this
line, the two phases discussed here are always present. These phases
are also delimited by a second line $s_1(\kappa)$ if the function $f$
admits at least one local extremum in the support of the typical density
of eigenvalues. These two lines intersect for each value of $\kappa$
such as $f'(\moy{\lambda_{\kappa N}}) = 0$, as illustrated in
Fig.~\ref{fig:Univ}.

Note that, away from the line $s_0(\kappa)$, other phases for the
Coulomb gas could emerge, depending on the choice of $f$.

\subsection{Outline of the paper}

The paper is organised as follows. In Section~\ref{sec:GenFormCoulGas}
we introduce the general formalism of the Coulomb gas, applied to the
study of truncated linear statistics. In Section~\ref{sec:fermions} we
analyse in details an application of this formalism to a specific
example of truncated linear statistics, motivated by the study of a
system of cold atoms. In Section~\ref{sec:Universality} we argue that
the main features observed on this example are actually universal, as
they neither depend on the choice of the matrix ensemble, nor on the
choice of truncated linear statistics under consideration.

%%%%%%%%%%%%%%%%%%%%%%%%%%%%%%%%%%%%%%%%%%%%%%%%%%%%%%%%%%%%%%%%%%%%%%%%%%%%%%%%%%%%%%%%%% 
%%%%%%%%%%%%%%%%%%%%%%%%%%%%%%%%%%%%%%%%%%%%%%%%%%%%%%%%%%%%%%%%%%%%%%%%%%%%%%%%%%%%%%%%%% 

\section{The Coulomb gas method for general truncated linear statistics}
\label{sec:GenFormCoulGas}

The idea of the Coulomb gas method is to rewrite the joint
distribution~(\ref{eq:DefInvEns}) as a Gibbs weight
\begin{equation}
  \label{eq:DefDiscrEner}
  \hspace{-2cm}
  P(\{ \lambda_n \}) \propto \e^{- \frac{\beta N^2}{2} E_{\mathrm{gas}}(\{ \lambda_n \})}
  \:,
  \quad
  E_{\mathrm{gas}}(\{ \lambda_n \}) =
    \frac{1}{N} \sum_{n=1}^N V(\lambda_n)
    - \frac{1}{N^2} \sum_{i \neq j} \ln \abs{\lambda_i-\lambda_j}
    \:.
\end{equation}
The energy $E_{\mathrm{gas}}$ describes a one dimensional gas of
particles at positions $\{ \lambda_n \}$, trapped in a confining
potential $V(\lambda)$ and submitted to repulsive logarithmic
interactions between each other. We have placed a factor $\beta$ in the
exponential in~(\ref{eq:DefInvEns}) so that this energy does not
depend on $\beta$.  In the limit $N \to \infty$, we expect that the
typical distribution of eigenvalues finds a balance between the
interaction and the confinement energy. This is achieved if
$\lambda_n = \O(N^0)$, which also implies that
$E_{\mathrm{gas}}(\{ \lambda_n \}) = \O(N^0)$. This is the reason why
we placed a factor $N$ in the definition~(\ref{eq:DefInvEns}): it
ensures that we manipulate quantities which do not scale with $N$. We
can then introduce the empirical density
\begin{equation}
  \rho(x) = \frac{1}{N} \sum_{n=1}^N \delta(x -\lambda_n)
  \:,
\end{equation}
which leads us to rewrite the measure~(\ref{eq:DefInvEns}) as (we
neglect the subleading entropic contributions, which are of order
$\O(N^{-1})$ \cite{Dys62,DeaMaj08})
\begin{equation}
  P(\lambda_1, \ldots, \lambda_N) \, \dd \lambda_1 \cdots \dd \lambda_N
  \rightarrow
  \e^{- \frac{\beta N^2}{2}  \mathscr{E}[\rho]} \: \D \rho
  \:,
\end{equation}
where the energy $\mathscr{E}[\rho]$ is the continuous version
of~(\ref{eq:DefDiscrEner}),
\begin{equation}
  \label{eq:DefEnergy}
  \mathscr{E}[\rho] = - \int \rho(x) \rho(y) \ln \abs{x-y} \dd x \dd y
  + \int \rho(x) V(x) \dd x 
  \:.
\end{equation}
Finally, we rescale the truncated linear
statistics~(\ref{def:TruncLinStat}) as
\begin{equation}
  \label{eq:defS}
  s = \frac{L_K}{N} = \int_c \rho(x) f(x) \dd x
  \:,
\end{equation}
where $c = \lambda_K$ is a lower bound ensuring that the summation
runs only on the $K$ largest eigenvalues. It can be determined as
\begin{equation}
  \label{eq:defCk}
  \int_c \rho(x) \dd x = \frac{K}{N} = \kappa  
  \:.
\end{equation}

Our aim is to compute the distribution of the rescaled truncated
linear statistics $s$, which can be expressed in terms of integrals
over the density:
\begin{eqnarray}
  \label{eq:PathIntegral}
  & \Pnk(s) 
  =
  \\\nonumber
  &\frac{\displaystyle \int \hspace{-0.1cm}\dd c \hspace{-0.1cm}\int\hspace{-0.1cm} \D\rho\:
    \e^{- \frac{\beta N^2}{2}  \mathscr{E}[\rho]} \:
    \delta\! \left(\int_c \hspace{-0.1cm} \rho(x) \dd x  - \kappa\right)
    \delta\! \left(\int \hspace{-0.1cm}\rho(x)\dd x - 1 \right)
    \delta\!\left(\int_c\hspace{-0.1cm} f(x) \rho(x) \dd x - s\right)}
    {\displaystyle \int\hspace{-0.1cm} \dd c \int\mathcal{D}\rho\:
    \e^{- \frac{\beta N^2}{2}  \mathscr{E}[\rho]} \:
    \delta\! \left(\int_c\hspace{-0.1cm} \rho(x) \dd x  - \kappa\right)
    \delta\! \left(\int\hspace{-0.1cm} \rho(x) \dd x  - 1 \right)}
  % \:.
\end{eqnarray}

%%%%%%%%%%%%%%%%%%%%%%%%%%%%%%%%%%%%%%%%%%%%%%%%%%%%%%%%%%%%%%%%%%%%%%%%%%%%%%%%%%%%%%%%%% 

\subsection{Saddle point equations and large deviation function}

When $N \to \infty$, the integrals in~(\ref{eq:PathIntegral}) are
dominated by the minimum of the energy $\mathscr{E}[\rho]$ under the
constraints imposed by the Dirac $\delta$-functions. These constraints
can be enforced by introducing three Lagrange multipliers
$\mu_0^{(1)}$, $\mu_0^{(2)}$, $\mu_1$. We thus consider
\begin{eqnarray}
  \label{eq:FreeEnergy}
  \mathscr{F}[\rho;\mu_0,\tilde{\mu}_0,\mu_1] 
  &= \mathscr{E}[\rho]
  + \mu_0^{(1)} \left(\int^c \rho(x) \dd x - (1-\kappa) \right)
  \nonumber\\
  &+ \mu_0^{(2)} \left(\int_c \rho(x) \dd x - \kappa \right)
   + \mu_1 \left(\int_c f(x) \rho(x) \dd x - s\right)
  \:.
\end{eqnarray}
Let us first focus on the numerator in Eq.~(\ref{eq:PathIntegral}),
and denote $\rho_\star(x;\kappa,s)$ the density that dominates these
integrals. It can be obtained in two steps. First, we find the density
$\tilde{\rho}(x;\mu_0^{(1)},\mu_0^{(2)},\mu_1)$ solution of
\begin{equation}
  \left.\frac{\delta \mathscr{F}}{\delta \rho(x)}\right|_{\tilde{\rho}}  = 0
  \:,
\end{equation}
which yields explicitly the integral equation
\begin{equation}
  \label{eq:SteepestDescent}
  \hspace{-1cm}
  2 \int \tilde{\rho}(y;\mu_0^{(1)},\mu_0^{(2)},\mu_1) \ln \abs{x-y} \dd y = V(x) + 
  \left\lbrace
    \begin{array}{ll}
      \mu_0^{(1)} & \text{for } x < c
      \\[0.125cm]
       \mu_0^{(2)} + \mu_1 f(x) & \text{for } x > c
    \end{array}
  \right.
\end{equation}
which can be understood as the energy balance for the particle at
point $x$ between the confinement and the logarithmic repulsion.  The
Lagrange multipliers $\mu_0^{(1)}$ and $\mu_0^{(2)}$ can be
interpreted as chemical potentials fixing the fraction of particles
respectively below and above $c$. The effect of the constraint on $s$
is to add an additional external potential, proportional to $\mu_1$,
which acts only on the $K$ rightmost eigenvalues.

Then, we determine the values $\mu_0^{(1)\star}(\kappa,s)$,
$\mu_0^{(2)\star}(\kappa,s)$, $\mu_1^\star(\kappa,s)$ of the Lagrange
multipliers in terms of the parameters $\kappa$ and $s$ by imposing
the constraints:
\begin{equation}
  \label{eq:ConstrNorm}
  \int_c \tilde{\rho}(x;\mu_0^{(1)\star},\mu_0^{(2)\star},\mu_1^\star) \dd x= \kappa
  \:,
  \quad
  \int^c \tilde{\rho}(x;\mu_0^{(1)\star},\mu_0^{(2)\star},\mu_1^\star) \dd x = 1-\kappa
\end{equation}
\begin{equation}
  \label{eq:ConstrTLS}
  \int_c f(x) \tilde{\rho}(x;\mu_0^{(1)\star},\mu_0^{(2)\star},\mu_1^\star) \dd x = s
  \:.
\end{equation}
Finally, the density which dominates the integrals in the numerator
of~(\ref{eq:PathIntegral}) is given by
\begin{equation}
  \label{eq:SolRhoStar}
  \rho_\star(x;\kappa,s) =  \tilde{\rho}(x;\mu_0^{(1)\star}(\kappa,s),\mu_0^{(2)\star}(\kappa,s),\mu_1^\star(\kappa,s))
  \:.
\end{equation}

For the denominator, we proceed similarly, but without the constraint
on $s$. The solution can be deduced from the one obtained above by
setting $\mu_1 = 0$. Explicitly, from the
solution~(\ref{eq:SolRhoStar}), it can be obtained by finding the
value $s_0(\kappa)$ which verifies $\mu_1(\kappa, s_0(\kappa)) =
0$. We then deduce the density which dominates the denominator
of~(\ref{eq:PathIntegral}) as
\begin{equation}
  \rho_0(x) = \rho_\star(x;\kappa,s_0(\kappa))
  \:.
\end{equation}

Having obtained the densities of eigenvalues $\rho_\star(x;\kappa,s)$
and $\rho_0(x)$ which dominate respectively the numerator and the
denominator of~(\ref{eq:PathIntegral}), we can evaluate the integrals
with a saddle point estimate, which yields
\begin{equation}
  \Pnk(s) \underset{N \to \infty}{\sim}
  \exp \left\lbrace
    - \frac{\beta N^2}{2} \Phi_\kappa(s)
  \right\rbrace
  \:,
\end{equation}
where we have introduced the large deviation function
\begin{equation}
  \Phi_\kappa(s) = \mathscr{E}[\rho_\star(x;\kappa,s)] - \mathscr{E}[\rho_0(x)]
  \:.
\end{equation}
This is the difference of energy between the two optimal
configurations of eigenvalues dominating the numerator and the
denominator of~(\ref{eq:PathIntegral}), respectively.  These energies
can be computed from the exact expressions of the densities using
Eq.~(\ref{eq:DefEnergy}), but this is in general a difficult
task. However, an important simplification was introduced in
Ref.~\cite{GraTex15}, based on the ``thermodynamic'' identity
\begin{equation}
  \frac{\dd \mathscr{E}[\rho_\star(x;\kappa,s)]}{\dd s} = - \mu_1^\star(\kappa,s)
  \:.
  \label{eq:thermoId}
\end{equation}
See Refs.\cite{CunFacViv16,GraTex16} for a more detailed discussion
of this relation. It can be used to obtain the large deviation
function via a simple integration of the Lagrange multiplier
$\mu_1^\star(\kappa,s)$ (which needs to be computed anyway to determine
$\rho_\star(x;\kappa,s))$:
\begin{equation}
  \label{eq:thermoId2}
  \Phi_\kappa(s) = \int_s^{s_0(\kappa)} \mu_1^\star(\kappa,t) \, \dd t
  \:.
\end{equation}
We will make extensive use of this relation to study of the
distribution $\Pnk(s)$.

To avoid cumbersome notations, the dependence of the density on the
parameters $\kappa$ and $s$ will be implicit from now on. We will also
not distinguish the densities
$\tilde{\rho}(x;\mu_0^{(1)},\mu_0^{(2)},\mu_1)$ and
$\rho_\star(x;\kappa,s)$; both will be denoted $\rho_\star(x)$ in the
following.

Now that we have laid out the procedure to obtain the distribution
$\Pnk(s)$, the main remaining task is to find the solution of the
saddle-point equation~(\ref{eq:SteepestDescent}).

\subsection{Reformulation and solution of the saddle-point equation}

In order to solve the saddle-point
equation~(\ref{eq:SteepestDescent}), it is convenient to take its
derivative:
\begin{equation}
  2 \dashint \frac{\rho_\star(y)}{x-y} \dd y = V'(x) + 
  \left\lbrace
    \begin{array}{ll}
      0 & \text{for } x < c\\
      \mu_1 f'(x) & \text{for } x > c
    \end{array}
  \right.
  \label{eq:SteepestDescentD}
\end{equation}
where $\dashint$ denotes a Cauchy principal value integral.  This
equation can be interpreted as the force balance on the eigenvalue
located at position $x$. The constraint on $s$ then acts as an
additional force $-\mu_1 f'(x)$ acting on the $K$ rightmost
eigenvalues (the sign comes from the fact that the force is the
opposite of the derivative of the potential). It is convenient to
split the density $\rho_\star$ into two densities: $\rhoR$ describing
the $K$ rightmost eigenvalues under consideration, and $\rhoL$ for the
others (see Figure~\ref{fig:DensPhIaII}),
\begin{equation}
  \rhoR(x) = \frac{1}{N} \sum_{n=1}^{K} \delta(x-\lambda_n)
  \:,
  \qquad
  \rhoL(x) = \frac{1}{N} \sum_{n=K+1}^{N} \delta(x-\lambda_n)
  \:.
  \label{eq:DefDensities}
\end{equation}
Due to the confining potential, the eigenvalues remain in a bounded
region in space, hence the densities $\rhoL$ and $\rhoR$ have compact
supports. Let us denote $[a,b]$ the support of $\rhoL$, and $[c,d]$
the support of $\rhoR$, where $c$ is the boundary introduced
previously in Eqs.~(\ref{eq:defS},\ref{eq:defCk}), as shown in
Figure~\ref{fig:DensPhIaII}. Note that it is possible that the two
supports merge, so that $b=c$, as shown in Fig.~\ref{fig:DensPhIaII}
(right).  We rewrite Eq.~(\ref{eq:SteepestDescentD}) in terms of these
densities as
\begin{eqnarray}
  \label{eq:SDRho1}
  2 \int_a^b \frac{\rhoL(y)}{x-y} \dd y + 2 \dashint_c^d \frac{\rhoR(y)}{x-y} \dd y
  =  V'(x) + \mu_1 f'(x)
  \hspace{1cm} 
  &\text{for } x \in [c,d]
  \\
  \label{eq:SDRho2}
  2 \dashint_a^b \frac{\rhoL(y)}{x-y} \dd y + 2 \int_c^d \frac{\rhoR(y)}{x-y} \dd y 
  = V'(x) 
  &\text{for } x \in [a,b]
  \:.
\end{eqnarray}
In these two equations, the principal value is only required when $x$
belongs to the domain of the integral. Such principal value integral
equations can be solved using a theorem due to Tricomi\cite{Tri57},
which gives an explicit expression for the inversion of Cauchy
singular equations of the form
\begin{equation}
  \dashint \frac{\rho(y)}{x-y} \dd y = g(x)
  \:,
  \label{eq:Tri1}
\end{equation}
under the assumption that the solution has one single support
$[a,b]$~\cite{Tri57}:
\begin{equation}
  \rho(x) = \frac{1}{\pi \sqrt{(x-a)(b-x)}} \left\lbrace
    A + \dashint_a^b \frac{\dd t}{\pi} \frac{\sqrt{(t-a)(b-t)}}{t-x} g(t)
  \right\rbrace
  \:,
  \label{eq:Tri2}
\end{equation}
where $A = \int_a^b \rho(x) \dd x$ is a constant. In our case, solving
the coupled equations~(\ref{eq:SDRho1},\ref{eq:SDRho2}) requires to
perform a double iteration of this theorem, in order to first
determine $\rhoL$ and then $\rhoR$, as in
Refs.\cite{MajNadScaViv11,GraMajTex16}. This procedure is rather
cumbersome, but it yields explicit expressions for the densities
$\rhoL$ and $\rhoR$.

For the case of the Laguerre ensembles of random matrix theory (which
corresponds to a specific $V(x)$ given in Table~\ref{tab:Ensembles})
and for monotonous functions $f$, the derivation is performed in the
Appendix of Ref.\cite{GraMajTex16}.  Here we adapt this procedure for
the general situation. The first step is to use Tricomi's theorem to
solve~(\ref{eq:SDRho2}) for $\rhoL$, treating $\rhoR$ as a known
function. Assuming that $\rhoL$ has a compact support $[a,b]$, we
directly apply~(\ref{eq:Tri2}), with
\begin{equation}
  g(t) = \frac{1}{2} V'(t) - \int_c^d \frac{\rhoR(y)}{t-y} \dd y
\end{equation}
and $A = \int_a^b \rhoL = 1-\kappa$ from the
constraint~(\ref{eq:ConstrNorm}). After permuting the integrals, and
using~(\ref{eq:PrincValInt1}), we obtain
\begin{eqnarray}
  \nonumber
  \rhoL(x) = \frac{1}{\pi \sqrt{(x-a)(b-x)}} \Bigg\lbrace
  1
  &+ \frac{1}{2} \dashint_a^b \frac{\dd t}{\pi} V'(t) \frac{\sqrt{(t-a)(b-t)}}{t-x}
  \\
  &
  - \int_c^d \dd y \: \rhoR(y) \frac{\sqrt{(y-a)(y-b)}}{y-x}
    \Bigg\rbrace
    \:.
    \label{eq:RhoLInterm}
\end{eqnarray}
Using now this expression in the second saddle-point
equation~(\ref{eq:SDRho1}), combined with the
integral~(\ref{eq:PrincValInt2}), we obtain an equation for $\rhoR$
only, valid for $x \in [c,d]$:
\begin{eqnarray}
  \nonumber
  \dashint_c^d \dd y \frac{\rhoR(y)}{x-y} \sqrt{(y-a)(y-b)}
  =& \frac{1}{2} \sqrt{(x-a)(x-b)} (V'(x) + \mu_1 f'(x))
  \\
   &- 1 + \frac{1}{2} \int_a^b \frac{\dd t}{\pi} \frac{V'(t)}{x-t} \sqrt{(t-a)(b-t)}
     \:.
\end{eqnarray}
We can solve this second equation using again Tricomi's
theorem~(\ref{eq:Tri2}), assuming that $\rhoR$ has a compact support
$[c,d]$. This yields
\begin{eqnarray}
  \nonumber
  \hspace{-2.5cm}
  \rhoR(x) = \frac{1}{\pi \sqrt{(x-a)(x-b)(x-c)(d-x)}} \Bigg\lbrace
  - \frac{1}{2} \int_a^b \frac{\dd t}{\pi} \frac{V'(t)}{t-x} \sqrt{(t-a)(b-t)(c-t)(d-t)}
  \\
   + C + x +  \frac{1}{2} \dashint_c^d \frac{\dd t}{\pi} \frac{V'(t) + \mu_1 f'(t)}{t-x} \sqrt{(t-a)(t-b)(t-c)(d-t)}
  \Bigg\rbrace
  \:,
  \label{eq:RhoRGen}
\end{eqnarray}
where $C$ is a constant that combines the integration constant from
Tricomi's theorem and other terms arising from the evaluation of the
integrals. The expression for $\rhoL$ can be obtained by plugging this
result into~(\ref{eq:RhoLInterm}),
\begin{eqnarray}
  \nonumber
  \hspace{-2.5cm}
  \rhoL(x) = \frac{-1}{\pi \sqrt{(x-a)(b-x)(c-x)(d-x)}} \Bigg\lbrace
  - \frac{1}{2} \dashint_a^b \frac{\dd t}{\pi} \frac{V'(t)}{t-x} \sqrt{(t-a)(b-t)(c-t)(d-t)}
  \\
   + C + x +  \frac{1}{2} \int_c^d \frac{\dd t}{\pi} \frac{V'(t) + \mu_1 f'(t)}{t-x} \sqrt{(t-a)(t-b)(t-c)(d-t)}
  \Bigg\rbrace
  \:.
  \label{eq:RhoLGen}
\end{eqnarray}
The constant $C$ can then be determined in the following way. Since
$c$ corresponds to the value $\lambda_K$ of the $K^{\mathrm{th}}$
eigenvalue, it can freely fluctuate. Therefore, we do not expect that
the density of eigenvalues diverges as $\rho_x(x) \sim (x-c)^{-1/2}$
for $x \to c$, as this type of behaviour typically occurs near a hard
edge, which is a hard constraint on the eigenvalues (such as
$\lambda_n > 0$). Therefore, the bracket in~(\ref{eq:RhoRGen}) must
vanish for $x=c$. This determines the value of the constant $C$, which
we can now use to simplify the
expressions~(\ref{eq:RhoRGen},\ref{eq:RhoLGen}). We can actually
express the total density $\rho_\star = \rhoL \cup \rhoR$ in a compact
form:
\begin{eqnarray}
  \nonumber
  \hspace{-2.5cm}
  \rho_\star(x) = \frac{1}{2\pi} \sqrt{\frac{c-x}{(x-a)(b-x)(d-x)}}
  \Bigg\lbrace
  2 + \dashint_a^b \frac{\dd t}{\pi} \frac{V'(t)}{t-x} \sqrt{\frac{(t-a)(b-t)(d-t)}{c-t}}
  \\
  \hspace{3.5cm}+\dashint_c^d \frac{\dd t}{\pi} \frac{V'(t) + \mu_1 f'(t)}{t-x} \sqrt{\frac{(t-a)(t-b)(d-t)}{t-c}}
  \Bigg\rbrace
  \:,
  \label{eq:RhoGen}
\end{eqnarray}
where the principal value must be applied only when $x$ is in the
domain of integration. This gives the general solution of the
saddle-point equation~(\ref{eq:SteepestDescentD}), in any invariant
ensemble~(\ref{eq:DefInvEns}) and for any truncated linear statistics
$f$, under the assumptions that both $\rhoL$ and $\rhoR$ have a
compact support. This will be the case for the example discussed
below, but some situations might lead to more complex solutions, which
would require to iterate Tricomi's theorem again for each additional
compact support. The constants $a$, $b$ and $d$ in~(\ref{eq:RhoGen})
will be determined by the boundary conditions (such as vanishing of
the density at the edge), while $c$ and $\mu_1$ will be fixed by the
constraints~(\ref{eq:ConstrNorm},\ref{eq:ConstrTLS}) which become
\begin{equation}
  \int_c^d \rhoR(x) \dd x = \kappa
  \hspace{1cm} , \hspace{1cm}
  \int_c^d \rhoR(x) f(x) \dd x = s
  \:.
  \label{eq:Constraints}
\end{equation}
Note that we have already used that $\rhoL$ normalises to $1-\kappa$
in the derivation above, so only the condition on $\rhoR$ remains.

We will see below that the general solution~(\ref{eq:RhoGen}) gives
rise to two different types of solutions (one with $b < c$ and the
other for $b=c$), which we will interpret as different phases for the
Coulomb gas. Instead of discussing the meaning of these phases and
their implication for the distribution of the truncated linear
statistics~(\ref{def:TruncLinStat}) on these general expressions, we
will consider a concrete example. We will discuss the generality of
the results obtained on this example in
Section~\ref{sec:Universality}.

%%%%%%%%%%%%%%%%%%%%%%%%%%%%%%%%%%%%%%%%%%%%%%%%%%%%%%%%%%%%%%%%%%%%%%%%%%%%%%%%%%%%%%%%%% 
%%%%%%%%%%%%%%%%%%%%%%%%%%%%%%%%%%%%%%%%%%%%%%%%%%%%%%%%%%%%%%%%%%%%%%%%%%%%%%%%%%%%%%%%%%
\section{Application to a system of fermions}
\label{sec:fermions}

In order to illustrate our analysis of truncated linear statistics, we
will study in details a specific example which arises from the physics
of cold atoms. Consider a system of $N$ spinless fermions in one
dimension, confined by a potential $V(x)$, described by the
Hamiltonian
\begin{equation}
  \mathcal{H} = \sum_{i=1}^N
  \left( - \frac{\hbar^2}{2m} \frac{\partial^2}{\partial x_n^2} + \mathcal{V}(x_n) \right)
  \:.
\end{equation}
The ground state of this system can be expressed in terms of the
one-particle eigenfunctions $\psi_k$ as a Slater determinant
\begin{equation}
  \Psi_0(x_1,\ldots,x_N) = \frac{1}{N!} \det[\psi_i(x_j)]_{1 \leq i,j \leq N}
  \:.
\end{equation}
This allows to establish a connection, for specific choices of
confining potential $V(x)$, between the positions of the trapped
fermions and the eigenvalues of random matrices. For instance, for a
harmonic trap $\mathcal{V}(x) = \frac{1}{2}m\omega^2 x^2$, the joint
distribution of the positions of the fermions is given by
\begin{equation}
  \label{eq:jpdfFermions}
  \abs{\Psi_0(x_1,\ldots,x_N)}^2 \propto \prod_{i<j} (x_i-x_j)^2 \prod_{n=1}^N \e^{-m \omega x_n^2/\hbar}
  \:.
\end{equation}
Introducing
\begin{equation}
  \lambda_n = x_n \sqrt{\frac{m \omega}{N \hbar}}
  \:,
\end{equation}
the joint distribution of the positions~(\ref{eq:jpdfFermions})
reduces to the joint distribution of eigenvalues~(\ref{eq:DefInvEns})
for the Gaussian Unitary Ensemble, corresponding to
\begin{equation}
  V(\lambda) = \lambda^2
  \quad \text{and} \quad
  \beta = 2
  \:.
\end{equation}
This relation has been used to study various observables, such as the
number of particles in a given interval\cite{MarMajSchViv16}. For a
review, see\cite{DeaLeDMajSch19}. For higher dimensional systems or
systems at finite temperature, the connection with random matrices is
lost. One can nevertheless use determinantal point processes to study
systems of noninteracting fermions in these
cases\cite{DeaLDoMajSch15,DeLDoMajSch15,DeLDoMajSch16,DeaLeDMajSch19,GraMajSchTex18}. Here,
we will focus on the zero temperature case, in one dimension, where
the system is in its ground state. We can therefore treat the
positions of the fermions as eigenvalues of random matrices from the
Gaussian Unitary Ensemble.

As an example of observable, we consider the potential energy carried
by the $K$ rightmost fermions:
\begin{equation}
  \label{eq:TruncEp}
  E_P(K) = \sum_{n=1}^K \frac{1}{2}m \omega^2 x_n^2
  = \frac{N^2 \hbar \omega}{2}  s \:,
  \quad
  s = \frac{1}{N} \sum_{n=1}^K \lambda_n^2
  \:.
\end{equation}
For $K=N$, the distribution of this observable can be studied by
standard techniques, and one can show that it follows a Gamma
distribution\cite{GreMajSch17}
\begin{equation}
  \label{eq:DistrFullLS}
  P_{N,\kappa=1}(s) = \frac{N^{N^2}}{\Gamma \left( \frac{N^2}{2} \right)}
  s^{\frac{N^2}{2}-1} \: \e^{-N^2 s}
  \:.
\end{equation}
For $K<N$, the observable~(\ref{eq:TruncEp}) is a truncated linear
statistics~(\ref{def:TruncLinStat}) with $f(\lambda) = \lambda^2$. We
now focus on the study of the distribution of this observable in the
regime $K \to \infty$, $N \to \infty$ with $\kappa = K/N$
fixed. Although this observable has a physical meaning only when
$\beta = 2$, we will obtain its distribution for any $\beta$ since the
derivation does not depend on this parameter.

%%%%%%%%%%%%%%%%%%%%%%%%%%%%%%%%%%%%%%%%%%%%%%%%%%%%%%%%%%%%%%%%%%%%%%%%%%%%%%%%%%%%%%%%%% 

\subsection{Optimal density without constraint}

The first step is to obtain the optimal density of eigenvalues
$\rho_0(x)$ in the absence of constraint. It is the density that
dominates the denominator of~(\ref{eq:PathIntegral}). This density
verifies the saddle point equation (\ref{eq:SteepestDescentD}) with
$\mu_1 = 0$:
\begin{equation}
  2 \dashint \frac{\rho_0(y)}{x-y} \dd y = 1
  \:,
\end{equation}
which can be solved using Tricomi's theorem
(\ref{eq:Tri1},\ref{eq:Tri2}). The density is the celebrated
semicircle distribution~\cite{Meh04,For10,AkeBaiDiFra11}
\begin{equation}
  \rho_0(x) = \frac{1}{\pi} \sqrt{2-x^2}
  \:.
\end{equation}

This density, obtained from $\mu_1=0$, corresponds to the maximum of
the probability $\Pnk$, and therefore to the most probable value of
$s$, given by
\begin{equation}
  s_0(\kappa) = \int_{c_0}^4 \rho_0(x) f(x)  \dd x =
  \frac{1}{2\pi} \arccos \frac{c_0}{\sqrt{2}}
  + \frac{c_0(1-c_0^2)}{4\pi} \sqrt{2-c_0^2}
  \:,
  \label{eq:optsC}
\end{equation}
where $c_0$ is fixed by the fraction $\kappa$ of eigenvalues we consider,
\begin{equation}
  \kappa = \int_{c_0}^4 \rho_0(x)  \dd x =
  \frac{1}{\pi} \arccos \frac{c_0}{\sqrt{2}}
  - \frac{c_0}{2\pi} \sqrt{2-c_0^2}
  \:.
  \label{eq:optkC}
\end{equation}
These two equations give a parametric representation of the line
$s_0(\kappa)$ in the $(\kappa,s)$ plane. It is the thick solid line
represented in Figure~\ref{fig:PhDiag}, which has the following
limiting behaviours
\begin{equation}
  s_0(\kappa)
  \simeq \left\lbrace
    \begin{array}{ll}
      \displaystyle
      2 \kappa % - \frac{2^{1/3} \pi^{3/2} (3\kappa)^{5/3}}{5}
      & \text{for } \kappa \to 0 \:,
      \\[0.3cm]
      \displaystyle
      \frac{1}{2} - 2 (1-\kappa)
      & \text{for } \kappa \to 1 \:.
    \end{array}
  \right.
\end{equation}

We will see in the following that $s_0(\kappa)$ defines a phase transition line.

%%%%%%%%%%%%%%%%%%%%%%%%%%%%%%%%%%%%%%%%%%%%%%%%%%%%%%%%%%%%%%%%%%%%%%%%%%%%%%%%%%%%%%%%%% 
\subsection{Phase I: two disjoint supports}
\label{sec:PhI}

We now turn to the general situation $\mu_1 \neq 0$, for which the
solution of the saddle point equation is given
by~(\ref{eq:RhoGen}). We first consider the situation where the
density $\rho_\star$ has two disjoint supports, i.e. $b<c$. In this
case, the general equations of Section~\ref{sec:GenFormCoulGas} give
for the optimal density,
\begin{eqnarray}
  \label{eq:RhoPhI}
  \rho_\star(x) =
  &
  \mathrm{sign}(x-b) \frac{\sqrt{(x-a)(b-x)(c-x)(d-x)}}{\pi}
  \\
  & \times \dashint_{[a,b]\cup [c,d]} \frac{\dd t}{\pi} \frac{\mathrm{sign}(t-b)}{t-x}
    \frac{t (1 + \mu_1 \Theta(t-c))}{\sqrt{(t-a)(t-b)(t-c)(d-t)}}
    \:,
\end{eqnarray}
where $\Theta$ is the Heaviside step function. We also have the
conditions coming from the vanishing of the density at $x=a$, $x=d$
and $x=b$,
\begin{equation}
  \label{eq:AphI}
  1 + \dashint_{[a,b]\cup[c,d]} \frac{\dd t}{\pi} t (1+\mu_1 \Theta(t-c))
  \sqrt{\frac{(d-t)(t-b)}{(t-a)(t-c)}}
  = 0
  \:,
\end{equation}
\begin{equation}
  \label{eq:DphI}
  \dashint_{[a,b]\cup[c,d]} \frac{\dd t}{\pi} t (1+\mu_1 \Theta(t-c))
  \sqrt{\frac{t-b}{(t-a)(t-c)(d-t)}}
  = 0
  \:,
\end{equation}
\begin{equation}
  \label{eq:BphI}
  \dashint_{[a,b]\cup[c,d]} \frac{\dd t}{\pi} \mathrm{sign}(t-c)
  \frac{t (1+\mu_1 \Theta(t-c))}
  {\sqrt{(t-a)(t-b)(t-c)(d-t)}}
  = 0
  \:,
\end{equation}
and the constraints~(\ref{eq:Constraints}). These expressions can be
written explicitly in terms of elliptic integrals, but they are more
compact in the integral form given above. The
density~(\ref{eq:RhoPhI}) is plotted in Fig.~\ref{fig:DensPhIaII}
(left).

This phase exists as long as the two supports remain disjoint, that is
$b<c$. We can actually obtain a necessary condition for this phase to
exist using a physical argument. We have indeed seen that the saddle
point equation~(\ref{eq:SteepestDescentD}) can be understood as a
force balance. The additional force in~(\ref{eq:SteepestDescentD})
acting on the $K$ rightmost eigenvalues is $-\mu_1 f'(x) = -\mu_1 x$,
where $x$ is the location of the eigenvalue subjected to this
force. Near the boundary $x=c$, this force is thus $-\mu_1 c$. For the
solution to have two supports, this force needs to be positive. This
gives the condition
\begin{equation}
  \label{eq:CondPhI}
  \mu_1 c < 0
\end{equation}
for the existence of Phase~I.

The limit of existence for this phase is $b=c$, when the two supports
merge into a single one. This will give rise to another phase of the
Coulomb gas, which we now study.

%%%%%%%%%%%%%%%%%%%%%%%%%%%%%%%%%%%%%%%%%%%%%%%%%%%%%%%%%%%%%%%%%%%%%%%%%%%%%%%%%%%%%%%%%% 
\subsection{Phase II: a logarithmic singularity}
\label{sec:PhII}

\begin{figure}
  \centering
  \includegraphics[width=\textwidth]{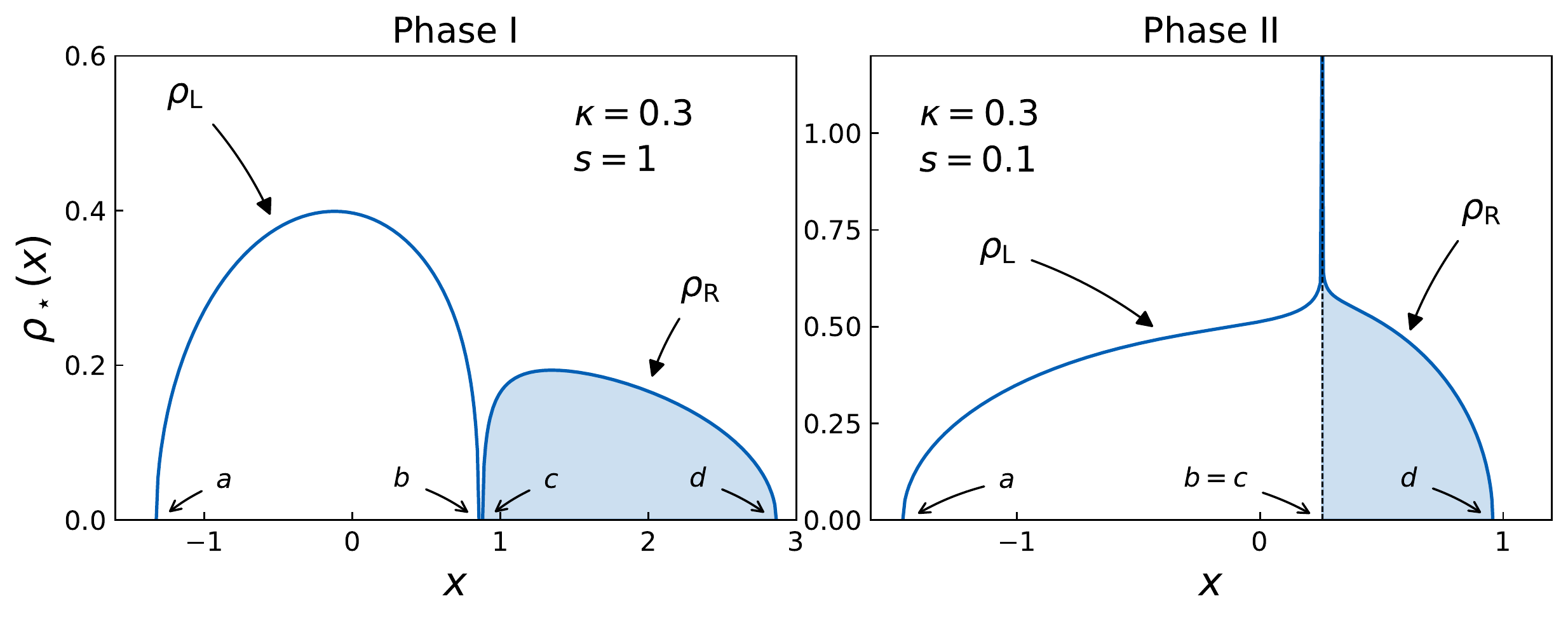}
  \caption{Optimal density of eigenvalues
    $\rho_\star=\rhoL \cup \rhoR$, solution of the saddle point
    equation~(\ref{eq:SteepestDescentD}), or equivalently
    Eqs.~(\ref{eq:SDRho1},\ref{eq:SDRho2}), for $\kappa=0.3$ and
    different values of $s$. Left: density for $s=1$, corresponding to
    Phase~I. Right: density for $s=0.1$, corresponding to Phase~II. In
    both cases, the fraction $\kappa$ of the largest eigenvalues are
    described by the density $\rhoR$, supported on $[c,d]$. The other
    eigenvalues are described by the density $\rhoL$, supported on
    $[a,b]$.}
  \label{fig:DensPhIaII}
\end{figure}

The second phase of interest here consists of a density with single
support $[a,d]$. It is deduced from the general
expression~(\ref{eq:RhoGen}) by setting $b=c$. The optimal density of
eigenvalues take a simpler form in this case:
\begin{eqnarray}
  \rho_\star(x) =
  &\frac{\sqrt{(d-x)(x-a)}}{\pi}
   \left(
    1 + \frac{2 \mu_1}{\pi} \arcsin \sqrt{\frac{d-c}{d-a}}
    \right)
    \nonumber
  \\
  &+ \frac{\mu_1 x}{\pi^2}
    \ln \abs{\frac{\sqrt{(x-a)(d-c)}+\sqrt{(c-a)(d-x)}}{\sqrt{(x-a)(d-c)}-\sqrt{(c-a)(d-x)}}}
    \:,
  \label{eq:RhoPhII}
\end{eqnarray}
along with the conditions coming from the vanishing of the density at $x=a$ and $x=d$,
\begin{eqnarray}
  \nonumber
   \hspace{-2.5cm} 1 &  \hspace{-2.3cm} + (1+\mu_1) \frac{(d-a)(3a+d)}{8}
  \\
  & \hspace{-2.3cm} +\frac{\mu_1}{4\pi} \left(
    (d-2c-3a)\sqrt{(c-a)(d-c)} - (d-a)(3a+d) \arcsin \sqrt{\frac{c-a}{d-a}}
  \right)
  = 0
    \:,
    \label{eq:VanishAPhII}
\end{eqnarray}
\begin{equation}
  \frac{a+d}{2} + \frac{\mu_1}{\pi}
  \left(
    \sqrt{(c-a)(d-c)} + (a+d) \arcsin \sqrt{\frac{d-c}{d-a}}
  \right)
  = 0
  \:,
    \label{eq:VanishDPhII}
\end{equation}
and the constraints~(\ref{eq:Constraints}). This gives four equations,
to determine the four parameters $a$, $c$, $d$ and $\mu_1$. The
density~(\ref{eq:RhoPhII}) is plotted in Fig.~\ref{fig:DensPhIaII}
(right). An important feature of the density~(\ref{eq:RhoPhII}) is
that is exhibits a logarithmic divergence at $x = c$:
\begin{equation}
  \label{eq:LogDivDens}
  \rho_\star(x) \underset{x \to c}{\simeq}
  - \frac{\mu_1 c}{\pi^2} \ln \abs{x-c}
  \quad \text{for} \quad c \neq 0
  \:.
\end{equation}
This behaviour had already been found in\cite{GraMajTex16} in the case
of a monotonous linear statistics. Additionally, when $c=0$, the
density no longer diverges, but presents a different logarithmic
singularity:
\begin{equation}
  \label{eq:LogBehavDens}
  \rho_\star(x) \underset{x \to c}{\simeq}
  \frac{\sqrt{-ad}}{\pi} - \frac{\mu_1 x}{\pi^2} \ln \abs{x}
  \quad \text{for} \quad c = 0
  \:.
\end{equation}
This type of singularity has, to the best of our knowledge, never been
found previously in the density of eigenvalues of random matrices. It
arises here because of the non-monotonicity of the function $f$ in the
truncated linear statistics~(\ref{def:TruncLinStat}).

\bigskip

A necessary condition for this phase to exist is that the
density~(\ref{eq:RhoPhII}) should remain positive for all
$x \in [a,d]$. In particular, from the
behaviour~(\ref{eq:LogDivDens}), this imposes that
\begin{equation}
  \label{eq:CondPhII}
  \mu_1 c > 0
  \:.
\end{equation}
This is the complementary condition of the one obtained for Phase~I,
see Eq.~(\ref{eq:CondPhI}). Since the expression~(\ref{eq:RhoPhII})
for Phase~II can be obtained by taking the limit $b \to c$ in the
general expressions of the density in Phase~I~(\ref{eq:RhoGen}), these
two phases should share a common boundary in the $(\kappa,s)$ plane,
which is thus given by $\mu_1 c = 0$. This gives two possibilities:
\begin{itemize}
\item $\mu_1 = 0$, which corresponds to the line $s=s_0(\kappa)$ in
  Fig.~\ref{fig:PhDiag}. This line was already present in the case of
  monotonous functions studied in Ref.\cite{GraMajTex16};
\item $c = 0$, which gives a new line $s = s_1(\kappa)$ in the phase
  diagram (Fig.~\ref{fig:PhDiag}). As we will discuss below, this line
  actually corresponds to the condition $f'(c) = 0$. The existence of
  this second line is thus specific to the study of truncated linear
  statistics with a non monotonous function $f$.

  Combining Eqs.~(\ref{eq:VanishAPhII},\ref{eq:VanishDPhII}) with the
  constraints~(\ref{eq:Constraints}), we can write this line $c=0$ in
  the parametric form
  \begin{equation}
    \label{eq:S1K1}
    \kappa = \frac{2}{\pi} \arccos \left( \sqrt{\frac{-a}{d-a}} \right)
    \:,
  \end{equation}
  \begin{equation}
    \label{eq:S1K2}
    s_1(\kappa) = (3 d^2 + 2 a d + 3 a^2 - 8) \frac{(-a d)^{3/2}}{16(a+d)\pi}
    \:,
  \end{equation}
  where $a$ and $d$ are related by
  \begin{equation}
    \label{eq:S1K3}
    \arccos \left( \sqrt{\frac{-a}{d-a}} \right)
    + (2 + ad)\frac{\sqrt{-a d}}{2(a+d)} = 0
    \:.
  \end{equation}
  It has the following asymptotic behaviour
  \begin{equation}
    s_1(\kappa)
    \simeq \left\lbrace
      \begin{array}{ll}
        \displaystyle
        \frac{2 \pi^4 \kappa^5}{45} % + \frac{44 \pi^6 \kappa^7}{4725}
        & \text{for } \kappa \to 0 \:,
        \\[0.3cm]
        \displaystyle
        \frac{3}{2\pi^2(1-\kappa)^3} % - \frac{1}{1-\kappa}
        & \text{for } \kappa \to 1 \:.
      \end{array}
    \right.
  \end{equation}
\end{itemize}

Note that we can recover the condition~(\ref{eq:CondPhII}) by
reversing the physical argument given in Section~\ref{sec:PhI}: the
eigenvalues near $x=c$, for $x>c$, feel the force $-\mu_1 c$. If this
force is positive, it pushes the eigenvalues to the right causing the
opening of a gap. In order to reverse the situation and get an
accumulation of eigenvalues near $x=c$, as it is the case here,
this force must be negative. This condition
yields~(\ref{eq:CondPhII}).

%%%%%%%%%%%%%%%%%%%%%%%%%%%%%%%%%%%%%%%%%%%%%%%%%%%%%%%%%%%%%%%%%%%%%%%%%%%%%%%%%%%%%%%%%% 
\subsection{Two infinite order phase transitions}

We have seen that the two lines $s_0(\kappa)$ and $s_1(\kappa)$
delimit regions in the $(\kappa,s)$ plane in which the optimal density
of eigenvalues $\rho_\star(x;\kappa,s)$ takes different forms. We can
thus interpret these lines as phase transitions for the Coulomb
gas. We now turn to the analysis of the order of these transitions.

\subsubsection*{Line $s_0(\kappa)$}--- We first consider the line on
which $\mu_1=0$, corresponding to the most probable value taken by the
truncated linear statistics~(\ref{def:TruncLinStat}). On this line,
the typical density of eigenvalues is given by Wigner's semicircle
law. One can show that all the derivatives of the energy of the
Coulomb gas $\mathscr{E}[\rho_\star(x;\kappa,s)]$, and thus of the
large deviation function $\Phi_\kappa(s)$ are continuous on this line
(see~\ref{app:InfOrderTr}). However, this function is not analytic: it
possesses an essential singularity in Phase I (corresponding to
$s=s_0^+$ for $\kappa < \frac{1}{2}$ and $s=s_0^-$ for
$\kappa > \frac{1}{2}$):
\begin{equation}
  \label{eq:InfPhTrI}
  \Phi_\kappa(s_0(\kappa) + \varepsilon) - \Phi_\kappa(s_0(\kappa) - \varepsilon)
  = \O( \varepsilon \: \e^{\gamma_0(\kappa)/\varepsilon} )
  \quad \text{for} \quad
  \kappa \neq \frac{1}{2}
  \:,
\end{equation}
where $\gamma_0(\kappa)$ is a constant. Therefore, in the standard
terminology of statistical physics, it corresponds to an
\textit{infinite order} phase transition. This exact same transition
has been observed in\cite{GraMajTex16} for truncated linear statistics
associated with a monotonous function $f$. Here, we obtain exactly the
same behavious for all values of $\kappa \neq \frac{1}{2}$. Indeed,
for these values of $\kappa$, the typical position of the
$K^{\mathrm{th}}$ largest eigenvalue (the last to contribute to $s$)
is $\lambda_K \neq 0$, away from the point where $f$ has a
minimum. Therefore, small fluctuations of $\lambda_K$ do not probe the
non-monotonicity of $f$, and the behaviour of $\Phi_\kappa$ near
$s_0(\kappa)$ is identical to the one observed in the monotonous case,
which has been shown to be universal\cite{GraMajTex16}.

\subsubsection*{Line $s_1(\kappa)$}--- The second line, corresponding
to $f'(c) = 0$ (that is, $c=0$ here, corresponding to
$\kappa = \frac{1}{2}$) is specific to the study of the case of
non-monotonous functions $f$. It is the main novelty that arises in this
case.

On this line, the density presents a logarithmis singularity, as shown
in Eq.~(\ref{eq:LogBehavDens}). It is a new specific feature to the
case of truncated linear statistics with a non-monotonous function
$f$.

One can show (see~\ref{app:InfOrderTr}) that the large deviation
function exhibits another essential singularity on this line (again
located in Phase I):
\begin{equation}
  \label{eq:InfPhTr1}
  \Phi_\kappa(s_1(\kappa) + \varepsilon) - \Phi_\kappa(s_1(\kappa) - \varepsilon)
  = \O( \varepsilon \: \e^{\gamma_1(\kappa)/\varepsilon} )
  \quad \text{for} \quad
  \kappa \neq \frac{1}{2}
  \:,
\end{equation}
where $\gamma_1(\kappa)$ is a constant. This shows that $s_1(\kappa)$
also corresponds to an \textit{infinite order} phase transition.

\subsubsection*{Intersection of the two lines for
  $\kappa = \frac{1}{2}$}--- The two phase transition lines intersect
for $\kappa = \frac{1}{2}$. Indeed, for this specific value of
$\kappa$, in the absence of constraint ($\mu_1 = 0$), the
$K^{\mathrm{th}}$ largest eigenvalue is typically located at
$\moy{\lambda_K} = c = 0$. At this point, the essential
singularities vanish, as well as the phase transition. Indeed, only
Phase I exists for this specific value of $\kappa$.

%%%%%%%%%%%%%%%%%%%%%%%%%%%%%%%%%%%%%%%%%%%%%%%%%%%%%%%%%%%%%%%%%%%%%%%%%%%%%%%%%%%%%%%%%% 
\subsection{Distribution of the potential energy of the $K$ rightmost fermions}

Using the results above on the optimal density $\rho_\star(x)$, we can
study the distribution of the truncated linear statistics under
consideration: the potential energy of the $K$ rightmost fermions in a
harmonic trap.

\subsubsection*{First cumulants}
\begin{figure}
  \centering
  \includegraphics[width=0.6\textwidth]{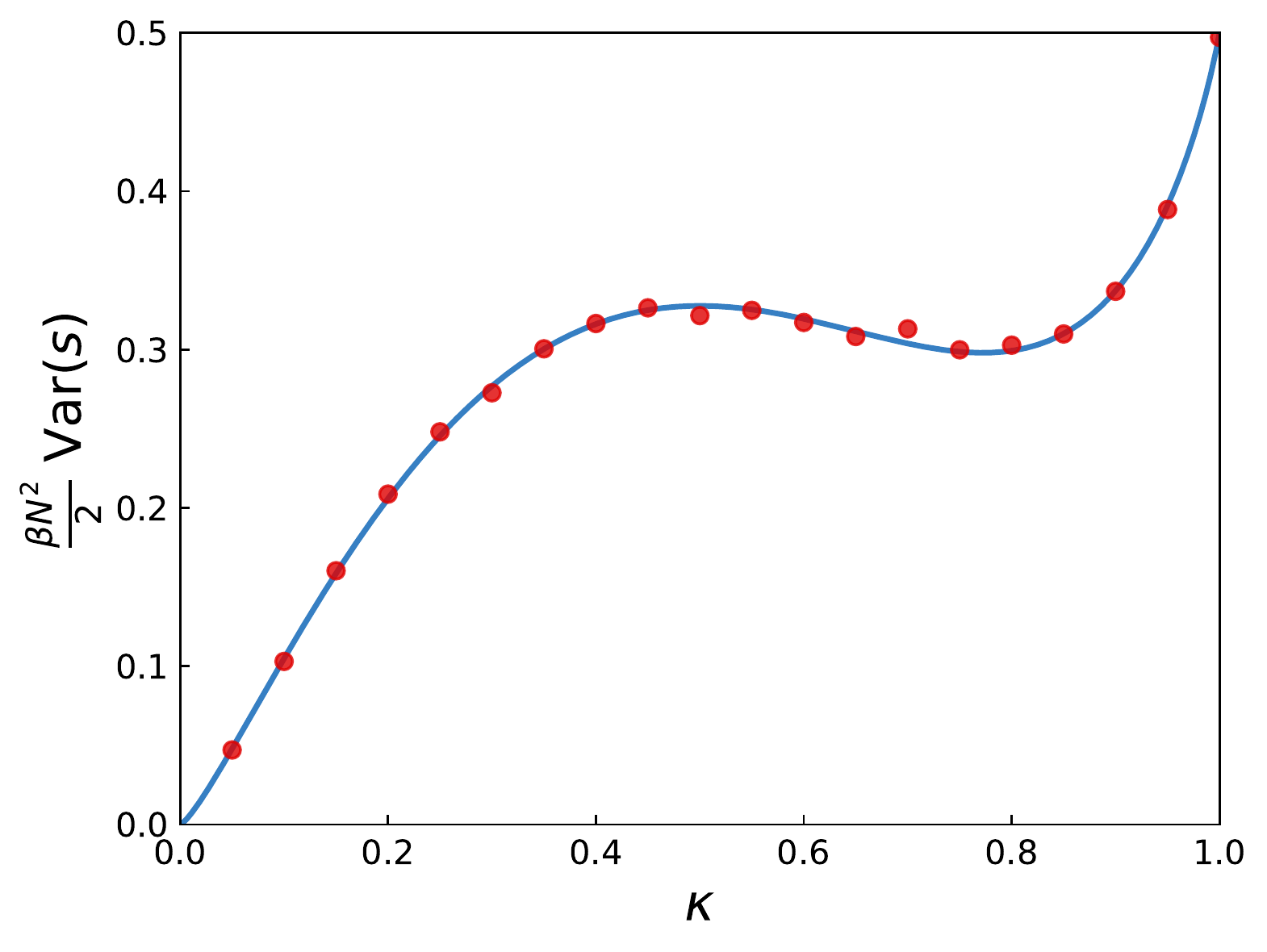}
  \caption{Variance of the rescaled truncated linear
    statistics~(\ref{eq:TruncEp}) as a function of $\kappa=K/N$,
    obtained from the parametric
    form~(\ref{eq:optkC},\ref{eq:ExprFc0},\ref{eq:ExprVarS}) (solid
    line). The points are obtained from the numerical computation of
    the variance of~(\ref{def:TruncLinStat}) by averaging over $20000$
    realisations of $200 \times 200$ matrices in the Gaussian Unitary
    Ensemble ($\beta=2$).}
  \label{fig:PltVar}
\end{figure}

--- The value $s_0(\kappa)$ corresponds to the most probable value taken by the truncated linear statistics, or equivalently by the potential energy of the $K$ rightmost fermions. It implies that
\begin{equation}
  \moy{E_P(K)} \underset{N,K \to \infty} \simeq \frac{N^2 \hbar \omega}{2} s_0(\kappa)
  \simeq \frac{\hbar \omega}{2} \left\lbrace
    \begin{array}{ll}
      \displaystyle
      2 K N & \text{for } K \to 0 \:,
      \\[0.4cm]
      \displaystyle
      2 K N - \frac{3}{2}N^2 & \text{for } K \to N \:.
    \end{array}
  \right.
\end{equation}
We can study the fluctuations around this value by expanding
Eqs.~(\ref{eq:Constraints},\ref{eq:VanishAPhII},\ref{eq:VanishDPhII})
in the limit $\mu_1 \to 0$. We obtain
\begin{equation}
  \label{eq:ExpSMu0}
  s = s_0(\kappa) + F(c_0) \mu_1 + \O(\mu_1^2)
  \:,
\end{equation}
with
\begin{equation}
  \label{eq:ExprFc0}
  \hspace{-2.5cm}
  F(c_0) =
  \frac{1}{8 \pi ^2} \left(3 c_0^4-14 c_0^2-4 \sqrt{2-c_0^2} \left(c_0^2-1\right) c_0
    \arccos\left(\frac{c_0}{\sqrt{2}}\right)+4 \arccos\left(\frac{c_0}{\sqrt{2}}\right)^2+16 \right)
\end{equation}
and $c_0$ is related to $\kappa$ via~(\ref{eq:optkC}). Inverting the
series~(\ref{eq:ExpSMu0}), we deduce the expression of the large
deviation function near $s_0(\kappa)$ via direct integration over $s$
thanks to the thermodynamic identity~(\ref{eq:thermoId2}):
\begin{equation}
  \Phi_\kappa(s) \underset{s \to s_0(\kappa)}{\simeq}
  \frac{(s-s_0(\kappa))^2}{2 F(c_0)}
  + \O( (s-s_0(\kappa))^3 )
  \:.
\end{equation}
From this result, we straightforwardly deduce
\begin{equation}
  \label{eq:ExprVarS}
  \mathrm{Var}(s) = \frac{2}{\beta N^2} F(c_0)
  \simeq \frac{2}{\beta N^2}
  \left\lbrace
    \begin{array}{ll}
      \displaystyle
      \left( \frac{18}{\pi} \right)^{2/3} \kappa^{4/3}
      & \text{for } \kappa \to 0
        \:,
      \\[0.4cm]
      \displaystyle
      \frac{1}{2} - 4 (1-\kappa) & \text{for } \kappa \to 1 \:.
    \end{array}
  \right.
\end{equation}
This variance is represented as a function of $\kappa$ in
Fig.~\ref{fig:PltVar}. It displays a non-monotonic behaviour, which
has never been observed in previous studies on truncated linear
statistics\cite{GraMajTex16,GraMajTex17,FlaMajSch21}. This new feature
has been confirmed by numerical simulations (see
Fig.~\ref{fig:PltVar}).

\subsubsection*{Behaviour for $s\to 0$}--- Looking at the phase
diagram in Fig.~\ref{fig:PhDiag}, we see that the limit $s \to 0$ is
reached in Phase I. We thus start from the expressions obtained in
this case. From the expression of the truncated linear
statistics~(\ref{eq:TruncEp}), we see that this limit corresponds to
$\lambda_n \to 0$ for all $n \leq K$. Therefore, for the optimal
density $\rho_\star$ it corresponds to $c\to 0$ and $d\to 0$. In order
to get the leading behaviour of the large deviation function, we only
need to expand Eq.~(\ref{eq:BphI}), which imposes $c=-d$, and then
expand the constraints~(\ref{eq:Constraints}), which give
\begin{equation}
  \kappa \simeq \mu_1 \frac{d^2}{2}
  \:,
  \qquad
  s \simeq \mu_1 \frac{d^4}{8}
  \:.
\end{equation}
We straightforwardly deduce that
\begin{equation}
  \mu_1 \underset{s \to 0}{\simeq} \frac{\kappa^2}{2s} + \O(1)
  \:,
\end{equation}
which yields the behaviour
\begin{equation}
  \Phi_\kappa(s) \underset{s \to 0}{\simeq}
  - \frac{\kappa^2}{2} \ln s + \O(1)
  \:,
\end{equation}
by direct integration, thanks to the thermodynamic identity~(\ref{eq:thermoId}).

\subsubsection*{Behaviour for $s\to +\infty$}--- From the phase
diagram in Fig.~\ref{fig:PhDiag}, the limit $s \to \infty$ is also
reached in Phase~I. But this time, it corresponds to making the
eigenvalues $\lambda_n$ for $n \leq K$ large. For the density
$\rho_\star$, it corresponds to $c,d \to \infty$. Expanding
Eqs.~(\ref{eq:AphI},\ref{eq:DphI},\ref{eq:BphI}), as well as the
constraints~(\ref{eq:Constraints}), we obtain:
\begin{eqnarray}
  1 + \frac{b-a}{8}(3a+b) \sqrt{\frac{c}{d}}
  + (1+\mu_1) \frac{d-c}{8}(3c+d)
  = 0
  \:,
  \\
  \frac{b-a}{8} \frac{3a+b}{\sqrt{c d}}
  + (1+\mu_1) \frac{c+d}{2} = 0
  \:,
  \\
  b = -a
  \:,
  \\
  \kappa = (1+\mu_1) \frac{(d-c)^2}{8}
  \:,
  \\
  s = (1+\mu_1)\frac{(d-c)^2}{128} (5c^2+6cd + 5d^2)
  \:.
\end{eqnarray}
Combining these expressions, we deduce that
\begin{equation}
  \mu_1 = -1 + \frac{\kappa(2-\kappa)}{2s} + \O(s^{-2})
  \:,
\end{equation}
which yields the behaviour
\begin{equation}
  \Phi_\kappa(s) \underset{s \to +\infty}{\simeq}
  s - \frac{\kappa(2-\kappa)}{2} \ln s + \O(1)
  \:,
\end{equation}
for the large deviation function, after using again the thermodynamic
identity~(\ref{eq:thermoId}).

\bigskip

\begin{figure}
  \centering
  \includegraphics[width=0.7\textwidth]{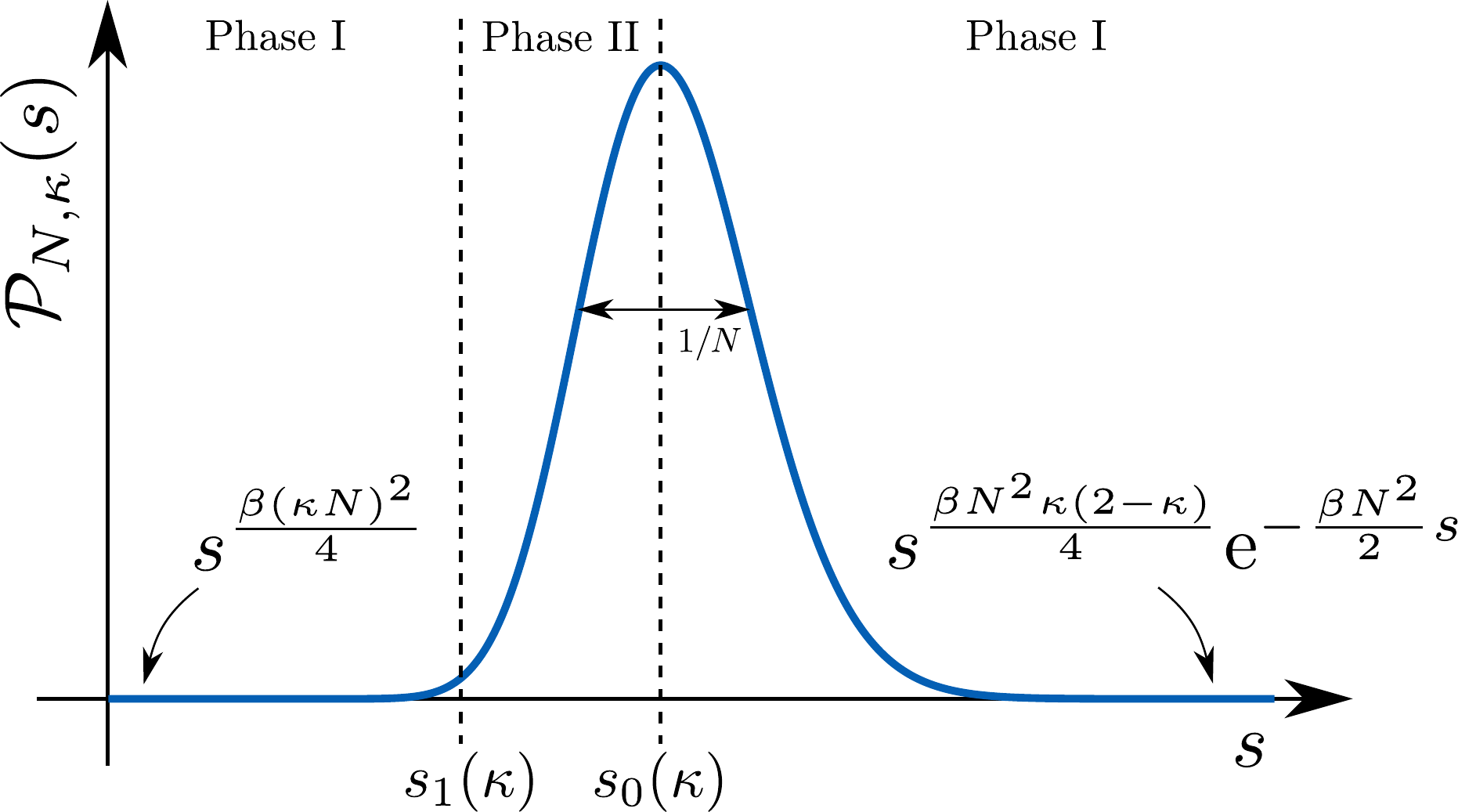}
  \caption{Sketch of the distribution of the truncated linear
    statistics~(\ref{def:TruncLinStat}), for $f(\lambda)=\lambda^2$ in
    the Gaussian ensembles. For $\beta=2$, it is the distribution of
    the potential energy~(\ref{eq:TruncEp}) of the $K = \kappa N$
    rightmost fermions in a harmonic trap. The large deviation
    function is not analytic, since it presents two essential
    singularities, located at $s=s_0(\kappa)$ and
    $s=s_1(\kappa)$. Here, these points are represented for
    $\kappa < \frac{1}{2}$. Otherwise, we have
    $s_1(\kappa) > s_0(\kappa)$, as shown in the phase diagram in
    Fig.~\ref{fig:PhDiag}.}
  \label{fig:SketchDistr}
\end{figure}

A sketch of the distribution of the truncated linear
statistics~(\ref{eq:TruncEp}) is shown in Fig.~\ref{fig:SketchDistr},
with all the behaviours identified in this Section. Unlike the case of
the monotonous linear statistics studied in Ref.\cite{GraMajTex16},
both asymptotic behaviours $s \to 0$ and $s \to \infty$ depend
explicitly on the fraction $\kappa$ of eigenvalues under
consideration. Indeed, in Ref.\cite{GraMajTex16}, these two limits
corresponded to two different phases of the Coulomb gas (equivalent to
Phase~I and Phase~II here). Here, they are both reached in Phase~I, in
which the $K$ eigenvalues of interest are separated from the others
and yield the dominating contribution to the energy of the Coulomb gas
(and thus to the large deviation function $\Phi_\kappa$).

Finally, for $\kappa=1$, we recover the known results for the full
linear statistics\cite{GreMajSch17}.

%%%%%%%%%%%%%%%%%%%%%%%%%%%%%%%%%%%%%%%%%%%%%%%%%%%%%%%%%%%%%%%%%%%%%%%%%%%%%%%%%%%%%%%%%% 
%%%%%%%%%%%%%%%%%%%%%%%%%%%%%%%%%%%%%%%%%%%%%%%%%%%%%%%%%%%%%%%%%%%%%%%%%%%%%%%%%%%%%%%%%%

\section{Universality}
\label{sec:Universality}

\begin{figure}
  \centering
  \includegraphics[width=0.7\textwidth]{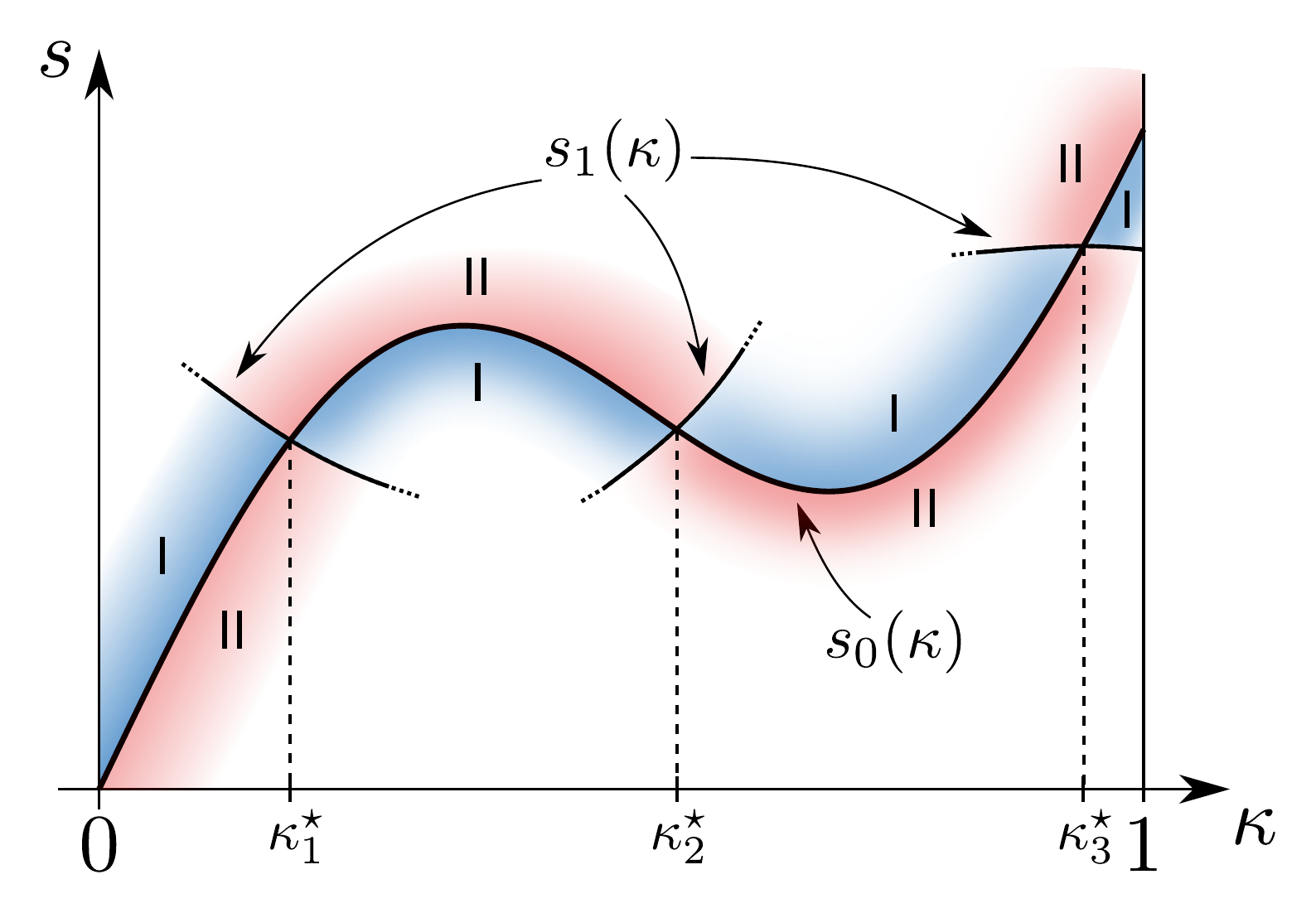}
  \caption{Sketch of the phase diagram for a truncated linear
    statistics for which $f$ has $3$ local extrema on the support of
    the typical density of eigenvalues $\rho_0$. The line
    $s_0(\kappa)$ corresponds to the typical value taken by the
    truncated linear statistics $s$, i.e. $\mu_1=0$. The second line
    $s_1(\kappa)$ is present only for non-monotonous functions $f$. It
    corresponds to $f'(c) = 0$. Note that Phase~I and Phase~II
    have been placed arbitrarily, and their precise positions depend
    on the specific choice of the function $f$.}
  \label{fig:Univ}
\end{figure}

In Section~\ref{sec:fermions} we have applied the Coulomb gas
formalism to the study of an example of truncated linear statistics
with a non monotonous function $f$, motivated by the study of a gas of
cold fermions. We now argue that several features of the Coulomb gas,
and thus of the large deviation function, are actually universal in
the sense that they do not depend on the specific choice of the linear
statistics (i.e. the function $f$) or on the matrix ensemble (i.e. the
potential $V$). Let us again denote $\rho_0$ the typical density of
eigenvalues in the ensemble under consideration, and
$c_0(\kappa) = \moy{\lambda_{\kappa N}}$ the position of the smallest
eigenvalue to contribute to the truncated linear
statistics~(\ref{def:TruncLinStat}). This position can be determined
from $\rho_0$ via the relation
\begin{equation}
  \kappa = \int_{c_0(\kappa)} \rho_0(x) \dd x
  \:.
\end{equation}
The following features are expected to occur in the study of any
truncated linear statistics.
\begin{itemize}
\item The line $s_0(\kappa)$, corresponding to $\mu_1=0$, gives the
  typical value of the truncated linear statistics restricted to the
  fraction $\kappa$ of the largest eigenvalues. It is shown
  in~\ref{app:InfOrderTr} that this line also corresponds to a
  transition line between two phases: a phase in which the density is
  supported on two disjoint supports (Phase~I), and one in which the
  density exhibits a logarithmic divergence (Phase~II). Moreover, it
  corresponds to an infinite order phase transition for all values of
  $\kappa$ such that $f'(c_0(\kappa)) \neq 0$. This extends the
  results of Ref.\cite{GraMajTex16} to the case of a non-monotonous
  function $f$.
\item The line $s_1(\kappa)$ is also present for any non-monotonous
  function $f$, at least in the vicinity of the typical line
  $s_0(\kappa)$. Indeed, away from this line, the two phases studied
  in this article could stop to exist and let other configurations of
  the Coulomb gas emerge (such as densities with more than two
  supports). Nevertheless, for values of $s$ close to $s_0(\kappa)$,
  the line $s_1(\kappa)$ is expected to exist, and is also an infinite
  order phase transition for the Coulomb gas, as is shown
  in~\ref{app:InfOrderTr}.
\item The two lines $s_0(\kappa)$ and $s_1(\kappa)$ intersect for
  values of $\kappa$ which verify $f'(c_0(\kappa)) = 0$. There are as
  many intersections as local extrema of the function $f$ in the
  support of $\rho_0$. At these points, there is no longer a phase
  transition in the Coulomb gas, as it remains in the same phase both
  for $s<s_0(\kappa)$ and $s>s_0(\kappa)$.
\item The positions of Phase~I and Phase~II with respect to the line
  $s_0(\kappa)$ are exchanged after a crossing with the line
  $s_1(\kappa)$.
\end{itemize}
We show in Fig.~\ref{fig:Univ} a sketch of the corresponding phase
diagram of the Coulomb gas, in the vicinity of the line $s_0(\kappa)$,
in the case where the lines $s_0(\kappa)$ and $s_1(\kappa)$ intersect
for $3$ different values of $\kappa$.

%%%%%%%%%%%%%%%%%%%%%%%%%%%%%%%%%%%%%%%%%%%%%%%%%%%%%%%%%%%%%%%%%%%%%%%%%%%%%%%%%%%%%%%%%% 
%%%%%%%%%%%%%%%%%%%%%%%%%%%%%%%%%%%%%%%%%%%%%%%%%%%%%%%%%%%%%%%%%%%%%%%%%%%%%%%%%%%%%%%%%% 

\section{Conclusion}
\label{sec:Conclusion}

We have studied the distribution of the truncated linear
statistics~(\ref{def:TruncLinStat}) with $f(\lambda)=\lambda^2$ in the
Gaussian ensembles (Eq.~(\ref{eq:DefInvEns}) with
$V(\lambda) = \lambda^2$) in the large $N$ limit, with
$\kappa = \frac{K}{N}$ fixed, using the Coulomb gas method. We have
shown that, for $\kappa \neq \frac{1}{2}$, the large deviation
function admits two essential singularities, at $s=s_0(\kappa)$ and
$s=s_1(\kappa)$, which result from two infinite order phase
transitions for the Coulomb gas. For the specific value
$\kappa = \frac{1}{2}$, these phase transitions merge and
disappear. We have argued that this picture is universal, and holds
for any choice of linear statistics $f$ or invariant matrix
ensemble~(\ref{eq:DefInvEns}), at least in the vicinity of the typical
value $s_0(\kappa)$ of the linear statistics.

As it was already noticed in Ref.\cite{GraMajTex16}, the limits
$\kappa \to 0$ and $\kappa \to 1$ are singular. Indeed, looking for
instance at $\kappa \to 1$ on the phase diagram in
Fig.~\ref{fig:PhDiag}, we would expect a phase transition at
$s = s_0(1) = \frac{1}{2}$, and thus an essential singularity in the
distribution of $s$ for $\kappa=1$. However, that is not the case as
this distribution is analytic, as easily seen on
Eq.~(\ref{eq:DistrFullLS}). This is due to the fact that the two
limits $\kappa \to 1$ and $N \to \infty$ do not commute. This
motivates a more detailed study of this joint limit, in order to
understand how the truncated linear statistics reduces to the full
linear statistics when $\kappa \to 1$.

The singularities observed on the large deviation function are
expected to be regularised by a smooth function when looking on a
scale $s-s_0(\kappa) = \O(N^{-\eta})$ with an exponent $\eta > 0$ (and
similarly for $s_1(\kappa)$). The most famous example is the case of
the largest eigenvalue: its large deviation function displays a
singularity which originates from a third order phase transition for
the Coulomb gas when looking at fluctuations of order
$\lambda_{\mathrm{\max}} - \moy{\lambda_{\mathrm{\max}}} = \O(1)$ (for
the distribution~(\ref{eq:DefInvEns}) for which $\lambda_n = \O(N^0)$
). This singularity is regularised when looking on a scale
$\Delta \lambda = \lambda_{\mathrm{\max}} - \moy{\lambda_{\mathrm{\max}}} =
\O(N^{-{2/3}})$, where $\Delta \lambda$ obeys the Tracy-Widom
distribution\cite{MajSch14}. We expect an equivalent function here
that regularises the distribution of the truncated linear statistics
at $s_0(\kappa)$ and $s_1(\kappa)$, when looking on a smaller
scale. It would be of particular interest to study the behaviour of
this function in vicinity of the point(s) where the two phase
transition lines $s_0(\kappa)$ and $s_1(\kappa)$ intersect, as there
is no singular behaviour at this point.  It could thus help to
understand how these essential singularities emerge in the
distribution of truncated linear statistics.

% \section*{Acknowledgements}

%%%%%%%%%%%%%%%%%%%%%%%%%%%%%%%%%%%%%%%%%%%%%%%%%%%%%%%%%%%%%%%%%%%%%%%%%%%%%%%%%%%%%%%%%% 
%%%%%%%%%%%%%%%%%%%%%%%%%%%%%%%%%%%%%%%%%%%%%%%%%%%%%%%%%%%%%%%%%%%%%%%%%%%%%%%%%%%%%%%%%% 
%%%%%%%%%%%%%%%%%%%%%%%%%%%%%%%%%%%%%%%%%%%%%%%%%%%%%%%%%%%%%%%%%%%%%%%%%%%%%%%%%%%%%%%%%% 

\appendix

%%%%%%%%%%%%%%%%%%%%%%%%%%%%%%%%%%%%%%%%%%%%%%%%%%%%%%%%%%%%%%%%%%%%%%%%%%%%%%%%%%%%%%%%%% 
%%%%%%%%%%%%%%%%%%%%%%%%%%%%%%%%%%%%%%%%%%%%%%%%%%%%%%%%%%%%%%%%%%%%%%%%%%%%%%%%%%%%%%%%%% 

\section{A few useful integrals}
\label{app:Integs}

The following integrals are used in the manuscript. 
\begin{equation}
  \label{eq:PrincValInt1}
  \hspace{-2cm}
  \dashint_a^b \frac{\dd t}{\pi} \frac{\sqrt{(t-a)(b-t)}}{t-x} \frac{1}{t-y}
  = -1 + \frac{\sqrt{(y-a)(y-b)}}{\abs{y-x}}
  \;,
  \quad
  x \in [a,b] \:,
  y \notin [a,b]
  \:,
\end{equation}
\begin{equation}
  \label{eq:PrincValInt2}
  \hspace{-1cm}
  \dashint_a^b \frac{\dd t}{\pi} \frac{1}{(x-t) \sqrt{(t-a)(b-t)}}
  =
  \left\lbrace
    \begin{array}{ll}
      \displaystyle - \frac{1}{\sqrt{(a-x)(b-x)}}  \:, & x < a \\[0.2cm]
      \displaystyle 0 \:, & a < x < b \\[0.2cm]
      \displaystyle \frac{1}{\sqrt{(x-a)(x-b)}} \:, & x > b \\[0.2cm]
    \end{array}
  \right.
\end{equation}

%%%%%%%%%%%%%%%%%%%%%%%%%%%%%%%%%%%%%%%%%%%%%%%%%%%%%%%%%%%%%%%%%%%%%%%%%%%%%%%%%%%%%%%%%% 
%%%%%%%%%%%%%%%%%%%%%%%%%%%%%%%%%%%%%%%%%%%%%%%%%%%%%%%%%%%%%%%%%%%%%%%%%%%%%%%%%%%%%%%%%% 

\section{Infinite order phase transitions}
\label{app:InfOrderTr}

We interpret the two lines $s_0(\kappa)$ and $s_1(\kappa)$ as phase
transitions lines for the Coulomb gas. We now show that these
transitions are of infinite order, for any invariant matrix ensemble
(choice of $V$) and any linear statistics ($f$). A similar analysis
was performed in Ref.\cite{GraMajTex16} in the Laguerre ensembles (see
Table~\ref{tab:Ensembles}), and for a monotonous function $f$. We
extend this discussion to the general case, and in particular to the
study of the line $s_1(\kappa)$ which emerges due to the non-monotony
of $f$.

Let us start from the general expression of the
density~(\ref{eq:RhoGen}). It depends on five parameters: $a$, $b$,
$c$, $d$ and $\mu_1$. These parameters are determined by the two
constraints~(\ref{eq:Constraints}), along with conditions on the
boundaries $a$, $b$ and $d$ of the support (the condition at $x=c$ has
already been used in the derivation of the
expression~(\ref{eq:RhoGen}) for the density). The determination of
the value of $b$ depends on the phase we consider:
\begin{itemize}
\item In Phase I in which the supports are disjoint ($b<c$), the value
  of $b$ is determined by imposing that the density does not diverge
  as $(b-x)^{-1/2}$ for $x \to b^-$ (as it was done for $c$), since
  this kind of behaviour only appears near a hard edge. This gives
  \begin{equation}
    \label{eq:GenCondB}
    \hspace{-2.5cm}
    2 = \int_a^b \frac{\dd t}{\pi} V'(t)
    \sqrt{\frac{(t-a)(d-t)}{(b-t)(c-t)}}
    -
    \int_c^d \frac{\dd t}{\pi} (V'(t) + \mu_1 f'(t))
    \sqrt{\frac{(t-a)(d-t)}{(t-b)(t-c)}}
    \:.
  \end{equation}
\item In Phase II, $b$ is directly obtained as $b=c$.
\end{itemize}

There remains to determine the outer edges $a$ and $d$. There are
different possibilities, depending on the random matrix ensemble under
consideration (and hence on the choice of $V$). In the main text,
since we worked in the Gaussian ensembles, we imposed that the density
vanishes at $x=a$ and $x=d$. In general, these conditions take the
form
\begin{equation}
  \label{eq:GenCdtVanishA}
  2 + \int_{[a,b]\cup[c,d]} \frac{\dd t}{\pi} (V'(t) + \mu_1 f'(t) \Theta(t-c))
  \sqrt{\frac{(b-t)(d-t)}{(t-a)(c-t)}} = 0
  \:,
\end{equation}
\begin{equation}
  \label{eq:GenCdtVanishD}
  2 - \int_{[a,b]\cup[c,d]} \frac{\dd t}{\pi} (V'(t) + \mu_1 f'(t) \Theta(t-c))
  \sqrt{\frac{(b-t)(t-a)}{(c-t)(d-t)}} = 0
  \:.
\end{equation}
Note that other types of boundary conditions can apply. For instance,
in the Jacobi ensembles in which the eigenvalues are restricted to
$[0,1]$ (see Table~\ref{tab:Ensembles}), we could have instead the
conditions $a=0$ and $d=1$. In the following, we will consider only
the case where the density vanishes at $x=a$ and $x=d$, and thus we
impose~(\ref{eq:GenCdtVanishA},\ref{eq:GenCdtVanishD}), since the
following procedure can be straightforwardly adapted to the other
types of boundary conditions.

\bigskip

In order to analyse the order of the two transitions occuring at
$s=s_0(\kappa)$ and $s=s_1(\kappa)$, we study the behaviour of
\begin{equation}
  \hspace{-1cm}
  \Phi_\kappa(s_0(\kappa) + \varepsilon)
  - \Phi_\kappa(s_0(\kappa) - \varepsilon)
  \quad \text{and} \quad
  \Phi_\kappa(s_1(\kappa) + \varepsilon)
  - \Phi_\kappa(s_1(\kappa) - \varepsilon)
\end{equation}
for $\varepsilon \to 0$. The two lines $s_0(\kappa)$ and $s_1(\kappa)$
corresponding to transitions between Phase I and Phase II, the limit
$\varepsilon \to 0$ in Phase I corresponds to the limit $b \to c$ in
both cases. Expanding the conditions on $a$, $b$ and
$d$~(\ref{eq:GenCondB},\ref{eq:GenCdtVanishA},\ref{eq:GenCdtVanishD}) in this limit,
we obtain for Phase~I:
\begin{equation}
  \mu_1 f'(c) = \frac{\alpha}{\ln(c-b)}
  \quad \Rightarrow \quad
  c-b = \e^{\alpha/(\mu_1 f'(c))}
  \:,
\end{equation}
\begin{equation}
  \label{eq:ConstAGen}
  2 + \int_a^d \frac{\dd t}{\pi} (V'(t) + \mu_1 f'(t) \Theta(t-c))
  \sqrt{\frac{d-t}{t-a}} =
  \O \left(\frac{\e^{\alpha/(\mu_1 f'(c))}}{\mu_1 f'(c)} \right)
  \:,
\end{equation}
\begin{equation}
  \label{eq:ConstDGen}
  2 - \int_a^d \frac{\dd t}{\pi} (V'(t) + \mu_1 f'(t) \Theta(t-c))
  \sqrt{\frac{t-a}{d-t}} =
  \O \left(\frac{\e^{\alpha/(\mu_1 f'(c))}}{\mu_1 f'(c)} \right)
  \:,
\end{equation}
where we have denoted
\begin{equation}
  \alpha = \frac{\pi}{\sqrt{(c-a)(d-c)}}
  \left(
    2 + \dashint_a^d \frac{\dd t}{\pi} \frac{V'(t)}{t-c}\sqrt{(t-a)(d-t)}
  \right)
  \:.
\end{equation}
Similarly, using the expression of the density~(\ref{eq:RhoGen}), the
constraints~(\ref{eq:Constraints}) take the form:
\begin{eqnarray}
  \nonumber
  \kappa = \int_c^d \frac{\dd x}{2\pi}
  &\frac{1}{\sqrt{(x-a)(d-x)}}
    \left\lbrace
    2 + \int_a^d \frac{\dd t}{\pi}
    \frac{V'(t)}{t-x} \sqrt{(t-a)(d-t)}
    \right.
  \\
  &+ \left.
    \mu_1 \int_c^d \frac{\dd t}{\pi} \frac{f'(t)}{t-x} \sqrt{(t-a)(d-t)}
    \right\rbrace
    + \O \left(\frac{\e^{\alpha/(\mu_1 f'(c))}}{\mu_1 f'(c)} \right)
    \:,
    \label{eq:ConstKGen}
\end{eqnarray}
\begin{eqnarray}
  \nonumber
  s = \int_c^d \frac{\dd x}{2\pi}
  &\frac{f(x)}{\sqrt{(x-a)(d-x)}}
    \left\lbrace
    2 + \int_a^d \frac{\dd t}{\pi}
    \frac{V'(t)}{t-x} \sqrt{(t-a)(d-t)}
    \right.
  \\
  &+ \left.
    \mu_1 \int_c^d \frac{\dd t}{\pi} \frac{f'(t)}{t-x} \sqrt{(t-a)(d-t)}
    \right\rbrace
    + \O \left(\frac{\e^{\alpha/(\mu_1 f'(c))}}{\mu_1 f'(c)} \right)
    \:.
    \label{eq:ConstSGen}
\end{eqnarray}
In Phase II, we obtain exactly the same expressions but without the
terms in $\O(e^{\alpha/(\mu_1f'(c))}/(\mu_1 f'(c)) )$. From these
expressions, we now consider each transition line independently.

\subsubsection*{Line $s_0(\kappa)$: $\mu_1 = 0$} ---
Let us first study the transition occuring on the line
$s_0(\kappa)$. It corresponds to $\mu_1=0$ and hence the most probable
value of the linear statistics. It is given in a parametric form by
\begin{equation}
  \label{eq:ParamS0KGen}
  s_0(\kappa) = \int_{c_0}^{d_0} f(x) \rho_0(x) \dd x
  \:,
  \quad
  \kappa = \int_{c_0}^{d_0} \rho_0(x) \dd x
\end{equation}
in terms of the position $c_0 = \moy{\lambda_K}$ of the last
eigenvalue to contribute to $s$, where we have denoted
\begin{equation}
  \rho_0(x) =  \frac{1}{\pi \sqrt{(x-a)(d-x)}}
  \left(
  1 + \dashint_a^d \frac{\dd t}{2\pi} \frac{V'(t)}{t-x} \sqrt{(t-a)(d-t)}
  \right)
\end{equation}
the typical density of eigenvalues in the absence of constraint.

For values of $c_0$ such that $f'(c_0) \neq 0$,
Eqs.~(\ref{eq:ConstAGen},\ref{eq:ConstDGen},\ref{eq:ConstKGen},\ref{eq:ConstSGen})
are identical to the ones studied in Ref.\cite{GraMajTex16} in the
case of a monotonous linear statistics. The fact that we recover these
exact same equations can be understood as follows: the small
fluctuations of $\lambda_K$ around the typical value $c_0$ are not
sufficient to probe the non-monotony of $f$ if $f'(c_0) \neq
0$. Therefore, for these values, the line $s_0(\kappa)$ corresponds to
an infinite order phase transition, as it was shown in
Ref.\cite{GraMajTex16}. The idea to prove this result is to expand
Eqs.~(\ref{eq:ConstAGen},\ref{eq:ConstDGen},\ref{eq:ConstKGen},\ref{eq:ConstSGen})
for $s \to s_0(\kappa)$, i.e. for $\mu_1 \to 0$, and combine them in
order to get
\begin{equation}
  s - s_0(\kappa) = F_0(\kappa) \mu_1 + \O( \mu_1^2 )
  + \O \left( \frac{\e^{\alpha/(\mu_1 f'(c_0))}}{\mu_1} \right)
  \quad \text{(Phase I)}
  \:,
\end{equation}
\begin{equation}
  s - s_0(\kappa) =  F_0(\kappa) \mu_1 + \O( \mu_1^2 )
  \quad \text{(Phase II)}
  \:.
\end{equation}
where $F_0(\kappa)$ is a constant fully determined by this
expansion. For the case considered in the main text
($f(x) = V(x) = x^2$), $F_0$ is given parametrically
by~(\ref{eq:optkC},\ref{eq:ExprFc0}). For the expansion in Phase I, we
have kept the subleading term
$\O \left( \frac{\e^{\alpha/(\mu_1 f'(c_0))}}{\mu_1} \right)$, as it
is the only one which differs from the expansion in Phase~II, since
all the terms in $\O(\mu_1^n)$ are identical in both
expressions. Inverting these expansions, we obtain $\mu_1$ as a
function of $s$, which can be integrated to yield $\Phi_\kappa(s)$ by
using the thermodynamic identity~(\ref{eq:thermoId2}). This gives
\begin{equation}
  \frac{\dd }{\dd \varepsilon}
  \left[
    \Phi_\kappa(s_0(\kappa) + \varepsilon) - \Phi_\kappa(s_0(\kappa) - \varepsilon)
  \right]
  = \O( \e^{\gamma_0(\kappa)/\varepsilon}/\varepsilon )
  \:,
\end{equation}
which after integration becomes
\begin{equation}
  \Phi_\kappa(s_0(\kappa) + \varepsilon) - \Phi_\kappa(s_0(\kappa) - \varepsilon)
  = \O( \varepsilon \: \e^{\gamma_0(\kappa)/\varepsilon} )
  \:,
\end{equation}
where we have denoted
\begin{equation}
  \label{eq:Gamma0}
  \gamma_0(\kappa) = \frac{\alpha}{f'(c_0)}F_0(\kappa)
  \:,
\end{equation}
with $c_0$ determined from $\kappa$ via~(\ref{eq:ParamS0KGen}). This
proves that all the derivatives of $\Phi_\kappa$ are continuous on the
line $s_0(\kappa)$, but the function is not analytic at this point,
due to the essential singularity present in Phase~I. However, for
specific values of $\kappa$ such that $f'(c_0)=0$ this singularity
vanishes and there is no transition. Indeed, the optimal density of
eigenvalues is given by the one of Phase~I in both cases
$s > s_0(\kappa)$ and $s < s_0(\kappa)$ (see for instance
Fig.~\ref{fig:PhDiag} for $\kappa = \frac{1}{2}$).

\subsubsection*{Line $s_1(\kappa)$: $f'(c) = 0$} --- We now turn to
the case of the second transition line, which is specific to the study
of non-monotonous truncated linear statistics. The behaviour of the
large deviation function for $s$ near $s_1(\kappa)$ can be obtained in
Phase I
from~(\ref{eq:ConstAGen},\ref{eq:ConstDGen},\ref{eq:ConstKGen},\ref{eq:ConstSGen})
by expanding these equations around $c = \check{c}$ such that
$f'(\check{c}) = 0$. For $c=\check{c}$, we have
$\mu_1 = \check{\mu}_1$. For $\check{\mu}_1 \neq 0$, we can combine
these expansions in order to write~(\ref{eq:ConstSGen}) as
\begin{equation}
  \hspace{-2cm}
  s - s_1(\kappa) = F_1(\kappa) (\mu_1 - \check{\mu}_1) + \O( (\mu_1 - \check{\mu}_1)^2 )
  + \O \left( \frac{\e^{\tilde{\gamma}_1(\kappa)/(\mu_1 - \check{\mu}_1)}}{\mu_1 - \check{\mu}_1} \right)
  \quad \text{(Phase I)}
  \:,
\end{equation}
where $F_1(\kappa)$ and $\tilde{\gamma}_1(\kappa)$ are two constants. We have kept
the subleading last term, as for Phase II, we have the same expansion,
but without this last term:
\begin{equation}
  s - s_1(\kappa) =
  F_1(\kappa) (\mu_1 - \check{\mu}_1) + \O( (\mu_1 - \check{\mu}_1)^2 )
  \quad \text{(Phase II)}
  \:.
\end{equation}
In these two expansions all the regular powers
$\O((\mu_1 - \check{\mu}_1)^n)$ are identical, the only difference is
the essential singularity present in Phase I only. The large
deviations function can be deduced from these expansions via the
thermodynamic identity~(\ref{eq:thermoId}), which yields
\begin{equation}
  \frac{\dd }{\dd \varepsilon}
  \left[
    \Phi_\kappa(s_1(\kappa) + \varepsilon) - \Phi_\kappa(s_1(\kappa) - \varepsilon)
  \right]
  = \O( \e^{\gamma_1(\kappa)/\varepsilon}/\varepsilon )
  \:,
\end{equation}
which after integration gives
\begin{equation}
  \Phi_\kappa(s_1(\kappa) + \varepsilon) - \Phi_\kappa(s_1(\kappa) - \varepsilon)
  = \O( \varepsilon \: \e^{\gamma_1(\kappa)/\varepsilon} )
  \:,
\end{equation}
where we have denoted
$\gamma_1(\kappa) = F_1(\kappa) \tilde{\gamma}_1(\kappa)$.  This
proves that the line $s_1(\kappa)$ is also associated to an infinite
order phase transition.

%%%%%%%%%%%%%%%%%%%%%%%%%%%%%%%%%%%%%%%%%%%%%%%%%%%%%%%%%%%%%%%%%%%%%%%%%%%%%%%%%%%%%%%%%% 
%%%%%%%%%%%%%%%%%%%%%%%%%%%%%%%%%%%%%%%%%%%%%%%%%%%%%%%%%%%%%%%%%%%%%%%%%%%%%%%%%%%%%%%%%% 

\end{document}